\documentclass[12pt,letterpaper,notitlepage]{article}
\pdfoutput=1

\usepackage{epsfig}
\usepackage{amsmath}
\usepackage{cite}
\usepackage{color}

\newcommand{\tev}{\,\, \mathrm{TeV}}
\newcommand{\gev}{\,\, \mathrm{GeV}}
\newcommand{\mev}{\,\, \mathrm{MeV}}
\newcommand{\lesim}{\,\raisebox{-.1ex}{$_{\textstyle<}\atop^{\textstyle\sim}$}\,}
\newcommand{\gesim}{\,\raisebox{-.1ex}{$_{\textstyle>}\atop^{\textstyle\sim}$}\,}

\begin{document}

\begin{titlepage}

\begin{flushright}
FERMILAB-PUB-14-038-T\\
PITT-PACC-1401
\end{flushright}

\vspace{15pt}
\begin{center}
  \LARGE Testing the Muon g--2 Anomaly at the LHC
\end{center}

\vspace{0pt}
\begin{center}
{\large A.~Freitas$^1$, J.~Lykken$^2$, S.~Kell$^1$, and S.~Westhoff$^1$}\\
\vspace{30pt} {
$^1$ PITTsburgh Particle-physics Astro-physics \& Cosmology
    Center (PITT-PACC),\newline  Department of Physics \& Astronomy,
    University of Pittsburgh, Pittsburgh, PA 15260, USA
   } \\
\vspace{10pt} {
$^2$ Theoretical Physics Department, Fermilab, P.O.\ Box 500, Batavia, IL 60510, USA
}
\end{center}

\vspace{10pt}
\begin{abstract}
\vspace{2pt} 
\noindent
The long-standing difference between the experimental measurement and
 the stan\-dard-model prediction for the muon's anomalous magnetic moment, $a_\mu = (g_{\mu}-2)/2$, may be explained by the presence of new weakly interacting particles with masses of a few \mbox{100 GeV.} Particles of this kind can generally be directly produced at the LHC, and thus they may already be constrained by existing data. In this work, we investigate this connection between $a_{\mu}$ and the LHC in a model-independent approach, by introducing one or two new fields beyond the standard model with spin and weak isospin up to one. For each case, we identify the preferred parameter space for explaining the discrepancy of $a_{\mu}$ and derive bounds using data from LEP and the 8-TeV LHC run. Furthermore, we estimate how these limits could be improved with the 14-TeV LHC. We find that the 8-TeV results already rule out a subset of our simplified models, while almost all viable scenarios can be tested conclusively with 14-TeV data.
\end{abstract}

\end{titlepage}

\clearpage


\section{Introduction}

The magnetic moment of the muon, $\vec{\mu}_\mu =
\frac{e}{2m_\mu}(1+a_\mu)\vec{\sigma}$, is one of the most precisely measured
quantities in particle physics and an important ingredient to electroweak
precision tests \cite{Beringer:1900zz}.\footnote{Here $\sigma_i$ ($i=1,2,3$) are the Pauli spin
matrices.} It is well known that the experimental value for the anomalous contribution $a_\mu$ from the Brookhaven E821
experiment \cite{exp} differs from the standard model (SM) prediction by about three
standard deviations. In particular, the analysis of \cite{review1} finds the discrepancy
\begin{equation}
\Delta a_\mu \equiv a_\mu^{\rm exp} - a_\mu^{\rm th} = (287 \pm 80) \times 10^{-11}.
\label{d}
\end{equation}
There are three generic possible sources for this discrepancy: \emph{(i)} the $a_\mu$ measurement itself, i.e.\ a statistical fluctuation or an 
overlooked systematic effect; \emph{(ii)} uncertainties in
the eva\-lua\-tion of non-perturbative hadronic corrections that enter in the SM prediction for $a_\mu$; or \emph{(iii)} loop corrections from new particles
beyond the SM. Concerning the first possibility, the experimental value will be
cross-checked by the E989 experiment at Fermilab \cite{Venanzoni:2012vha} and the planned $g{-}2$/EDM experiment at J-PARC \cite{Saito:2012zz} in the near future. 
The hadronic corrections are difficult to evaluate,
requiring input from experimental data, perturbative QCD, and non-perturbative
hadronic models. However, several recent evaluations \cite{amue} yield results
that all confirm a discrepancy of about $3\sigma$ or more.

In the presence of physics beyond the standard model (BSM), the leading one-loop
contribution is parametrically of the order of $\delta a_\mu \sim \frac{g^2_{\rm
NP}}{16\pi^2}\,\frac{m_\mu^2}{M^2_{\rm NP}}$, which can match the observed
discrepancy for ${\cal O}(1)$ values of the couplings, $g_{\rm NP}$, and
${\cal O}(100\gev)$ values of the masses, $M_{\rm NP}$, of the new particles. These
ingredients can be satisfied by a large number of new-physics models, such as
supersymmetry, extended gauge groups, extra dimensions, seesaw models, or extended
Higgs sectors (see \cite{review1} and references therein).

In this article, rather than studying concrete BSM
models and their impact on $a_\mu$, we analyze minimal sets of new particles
that can produce a one-loop correction of the required size. For
definiteness, we consider one or two new fields with different spins and
gauge-group representations. To allow a perturbative description for the $a_\mu$
correction, we focus on weakly coupled new physics, i.e. $|g_{\rm NP}| \lesim \sqrt{4\pi}$. We are interested in scenarios that can, at least in principle, be tested at
collider experiments. Thus we do not consider very light superweakly coupled
new particles, which can also successfully explain the $a_\mu$ discrepancy \cite{superweak}. Instead, we restrict ourselves to new particles
with weak-scale masses $M_{\rm NP}\gesim 100\gev$. Particles of this kind are generically within reach of
the LHC and may be additionally constrained by data from LEP.

The main goal of this paper is to establish a relationship between weak-scale
BSM explanations for the discrepancy of the muon anomalous magnetic moment and
direct searches for these particles at the LHC. After defining the overall
framework and generic constraints in Section~\ref{sec:ew}, we compute in
Sections~\ref{sec:onefield}--\ref{sec:twofields} the corrections to $a_\mu$ by
adding one new field, two new mixed fermion fields, and two
new fields with different spins to the SM, respectively. For each of these cases, we evaluate the
viable parameter space that can explain the discrepancy in \eqref{d}, given
constraints from LEP and other lower-energy experiments. In
Section~\ref{sec:lhc}, we explore how the viable new-physics scenarios can be probed at the
LHC by recasting exi\-sting new-physics searches published by the ATLAS and CMS
collaborations. While these experimental searches are generally not optimized for our purposes, they nevertheless lead to non-trivial constraints on new-physics explanations for the $a_\mu$ discrepancy. We also estimate how the reach could be extended with the full 14-TeV run of the LHC. In Section~\ref{sec:tanb}, we briefly comment on new-physics models where the
$a_\mu$ correction is enhanced by $\tan\beta$, the ratio of the vacuum
expectation values (vevs) of two Higgs doublets, which is not covered by the cases
discussed in the previous sections. Finally, the conclusions are presented in
Section~\ref{sec:concl}.


\section{Electroweak contributions}
\label{sec:ew}

Electroweak SM contributions to $a_{\mu}$ are suppressed by $\mathcal{O}(m_{\mu}^2/M_{W}^2) = 10^{-6}$ with respect to QED contributions, due to the exchange of the massive gauge bosons.\footnote{The contributions from Higgs bosons receive an additional suppression by $m_{\mu}^2/M_H^2$ from the muon Yukawa coupling.} 
At the one-loop level, they yield \cite{Beringer:1900zz}
\begin{equation}
a_{\mu}^{\text{EW}} = \frac{G_F m_{\mu}^2}{8\sqrt{2}\pi^2}\left[\frac{5}{3} + \frac{1}{3}(1-4\sin^2\theta_W)^2 + \mathcal{O}\left(\frac{m_{\mu}^2}{M_{\text{EW}}^2}\right)\right] = 194.8\times 10^{-11}\,,
\end{equation}
with the Weinberg angle $\sin^2\theta_W \approx 0.2236$ and Fermi constant $G_F = 1.16638\times 10^{-5}\gev^{-2}$. Generically, new weakly-coupled particles with electroweak-scale masses $M_{\text{EW}}$ will yield corrections of comparable size. Since the magnetic moment breaks parity, any contribution to $a_{\mu}$ involves a flip of the muon's chirality. This is typically achieved by a mass term, which breaks the chiral symmetry of the underlying theory. New electroweak contributions to $a_{\mu}$ are therefore expected to exhibit the same suppression $\mathcal{O}(m_{\mu}^2/M_{\text{EW}}^2)$ as in the SM.

We aim at performing a model-independent analysis of contributions to $a_{\mu}$ from new particles around the electroweak scale. We consider all possible one-loop contributions of fields with spin $0$, $1/2$ and $1$ that are singlets, doublets or triplets under the gauge group SU(2) of weak interactions, and with integer electric charges. In Table~\ref{tab:notation}, we introduce the corresponding notation and give examples of models which incorporate such new particles. Their contributions to $a_{\mu}$ can be classified with respect to the fields occurring in the loop:
\begin{enumerate}
\item One new field and a SM lepton, $W$, $Z$ or Higgs boson (Figure~\ref{fig:onep}).
\item Two new mixing fermions and a $W$, $Z$ or Higgs boson (Figure~\ref{fig:amu-meg-mmu}, left).
\item Two new fields with different spins (Figure~\ref{fig:twop}).
\end{enumerate}
We will discuss these three categories one by one in the following sections. Contributions with two mixing fermions (2.) always imply contributions with one new fermion (1.). All other two-field contributions (3.) may imply one-field contributions (1.). The latter, however, can be strongly constrained by measurements of other observables (as will be discussed in the following subsections) or entirely prohibited due to a discrete symmetry.\footnote{A prominent example for such a symmetry is $R$ parity in models with supersymmetry.} Diagrams with two new fields in the loop can therefore become the dominant contribution to $\Delta a_{\mu}$. In addition to contributions from new particles in the loop, the electroweak SM contributions to $a_{\mu}$ can be modified by the mixing of new fermions with SM leptons through corrections to the lepton gauge couplings and Yukawa couplings. In models that incorporate at least two scalar fields with vevs $v_1$ and $v_2$, additional contributions enhanced by $\tan\beta = v_1/v_2$ occur. These effects will be discussed separately in Section \ref{sec:tanb}.

\begin{table}
\centering
\renewcommand{\arraystretch}{1.5}
\begin{tabular}{|l|l|l|}
\hline
Vector bosons & $V^0,V^{\pm},V_A$ & $Z',W'$, left-right symmetric electroweak sector ($V_A$)\\
Scalar bosons & $\phi^0,\phi^{\pm},\phi_D,\phi_A,\phi_T$ & extended Higgs sectors, seesaw type II ($\phi_T$)\\
Fermions & $\psi^0,\psi^{\pm},\psi_D,\psi_A,\psi_T$ & composite fermions, seesaw type III ($\phi_A$)\\
\hline
\end{tabular}
\caption{New fields considered in this work, their electroweak properties and examples for models in which they appear. $0,\pm$: neutral, charged weak singlets. $D$: weak doublet with hypercharge $\pm 1/2$. $A,T$: weak triplets with hypercharge $0,-1$.}
\label{tab:notation}
\end{table}


\subsection{Constraints from LEP observables}
New electroweak contributions to $a_{\mu}$ are generally constrained by precision observables and direct searches at LEP. In this section, we study generic constraints on the masses and couplings of new particles that apply to all the cases discussed in the following sections. We focus on robust constraints with a model-independent connection to $a_{\mu}$. Along those lines, processes involving couplings to quarks are not taken into account, since they can easily be circumvented in hadrophobic models.

Direct mass constraints on new particles can be obtained from LEP II searches for pair production via gauge interactions with a $Z$ boson or photon, namely $e^+ e^-\rightarrow Z/\gamma\rightarrow XX$. Assuming one dominant decay mode (new bosons decay mainly into leptons, new fermions decay via electroweak currents through mixing with SM leptons), mass constraints are independent from the couplings to fermions. The non-observation of new vector bosons, scalars and fermions at center-of-mass (CM) energies around $\sqrt{s}\approx 200\gev$ yields a general mass bound of $M\gesim 100\gev$ (see for instance the listings for Higgs bosons, heavy charged-lepton searches, and other lepton searches in \cite{Beringer:1900zz}). These constraints do not apply to SM gauge singlets, which cannot be produced through electroweak interactions.\footnote{Since we assume that new scalar fields do not acquire a vev, associated production with a $Z$ boson is prohibited.}

The exchange of a new heavy scalar or gauge boson in $e^+e^-\rightarrow \ell^+\ell^-$ processes leads to four-lepton contact interactions, which are strongly constrained by LEP measurements. Details will be discussed in Section~\ref{sec:onefield}. Besides the resonant production of one new particle, similar constraints also apply to couplings of two new particles to a lepton, which generate four-lepton interactions at one-loop level. Due to the loop suppression, the bounds are generally weaker than for one new particle, but important if new particles couple strongly to leptons. One-loop effects on four-lepton interactions will be discussed in detail in Section~\ref{sec:twofields}, analytic results are given in Appendix~\ref{sec:4lci}. We emphasize that our results are model-independent and can thus be of general use to constrain the couplings of two new particles to leptons from LEP measurements.

Strong constraints on new particles in weak multiplets arise from the ``oblique'' para\-meters $S$ and $T$ \cite{Peskin:1991sw}. The $T$ parameter is sensitive to weak isospin breaking through mass splitting among the multiplet constituents. To prevent large contributions to $T$, we require (approximate) mass degeneracy for the components of SU(2) doublets or triplets. The $S$ parameter probes different isospin three-components $T_3$ of left- and right-chiral fermions, $S\sim \big[T_3(\psi_L)-T_3(\psi_R)\big]^2$. To avoid such effects, we impose vector-like couplings of new fermions to gauge bosons. This simultaneously ensures the cancellation of axial-vector gauge anomalies.

In summary, we assume the following properties of new particles in our analysis:
\begin{itemize}
\item Particles with electroweak quantum numbers are heavier than $100\gev$.
\item Constituents of weak multiplets are degenerate in mass.
\item Couplings involving new particles are real and perturbative, \emph{i.e.} smaller than $\sqrt{4\pi}$.
\item New fermions have vector-like electroweak couplings.
\item All interactions involving leptons are minimally flavor-violating.
\end{itemize}
By limiting ourselves to couplings without an imaginary part, we circumvent potential (model-dependent) constraints from the electric dipole moment of the electron \cite{Baron:2013eja}. The assumption of minimal flavor violation (MFV) is motivated by strong constraints from the process $\mu\rightarrow e\gamma$ and from the smallness of the muon mass. These constraints and their relation to $a_{\mu}$ will be discussed in detail in the following Section~\ref{sec:amu-meg-mmu}.


\subsection{Constraints from $\mathcal{B}(\mu \rightarrow e \gamma)$ and the muon mass}\label{sec:amu-meg-mmu}
The flavor-conserving anomalous magnetic moment $a_{\mu}$ is tightly connected to the flavor-violating process $\mu\rightarrow e\gamma$. In the framework of an effective theory, new-physics contributions to both quantities are described by dimension-six operators with the same gauge and Lorentz structure \cite{Cirigliano:2005ck},
\begin{align}
O^1_{a_{\mu}} &=  g'y_{\mu}H^{\dagger}\overline{\mu_R}\sigma^{\mu\nu}\mu_L B_{\mu\nu}, & & O^1_{\mu e} = g'y_{\mu} H^{\dagger}\overline{e_R}\sigma^{\mu\nu}\Delta_{\mu e}\mu_L B_{\mu\nu},\nonumber\\
O^2_{a_{\mu}} &= gy_{\mu} H^{\dagger}\overline{\mu_R}\sigma^{\mu\nu}\tau^a\mu_L W_{\mu\nu}^a, & & O^2_{\mu e} = gy_{\mu} H^{\dagger}\overline{e_R}\sigma^{\mu\nu}\Delta_{\mu e}\tau^a\mu_L W_{\mu\nu}^a,
\end{align}
where $y_\mu$ is the muon Yukawa coupling, $H$ is the SM Higgs doublet with vev $v=246\gev$, and $B_{\mu\nu}$ and $W_{\mu\nu}^a$ are the U(1) and SU(2) gauge fields before electroweak symmetry brea\-king with the corresponding gauge couplings $g'$ and $g$. The labels $L,R$ on the fermion fields denote left- and right-chiral states, respectively, while $\overline{\psi_{L,R}}$ denote anti-fermions with the same chirality, i.e.\ opposite helicity. The amount of flavor violation is parametrized by $\Delta_{\mu e}$. The branching ratio of $\mu\rightarrow e\gamma$ normalized to $\mu\rightarrow e \nu_{\mu}\overline{\nu}_e$ is given by \cite{Cirigliano:2005ck}
\begin{equation}
\mathcal{B}(\mu\rightarrow e\gamma) = 384\,\pi^2e^2\frac{v^4}{\Lambda_{\text{FV}}^4}|\Delta_{\mu e}|^2\big|C^1_{\mu e} - C^2_{\mu e}\big|^2\approx 6.34\times 10^{-7}\left(\frac{1\tev^4}{\Lambda_{\text{FV}}^4}\right)|\Delta_{\mu e}|^2\,,
\end{equation}
where $C^i_{\mu e}\approx \mathcal{O}(1)$ are Wilson coefficients and $\Lambda_{\text{FV}}\gg v$ denotes the scale at which lepton flavor violation occurs explicitly through new degrees of freedom. The current experimental bound $\mathcal{B}(\mu\rightarrow e\gamma) < 5.7\times 10^{-13}$ \cite{Adam:2013mnn} imposes strong constraints on $|\Delta_{\mu e}|/\Lambda_{\text{FV}}^2$. This implies that contributions to $a_{\mu}$ from a scale $\Lambda\approx\Lambda_{\text{FV}}\lesim 1\tev$ (necessary to explain the discrepancy $\Delta a_{\mu}$) are ruled out, unless a protection mechanism is at work that suppresses the flavor violation $\Delta_{\mu e}$.

\begin{figure}
\centering
\includegraphics[scale=0.65]{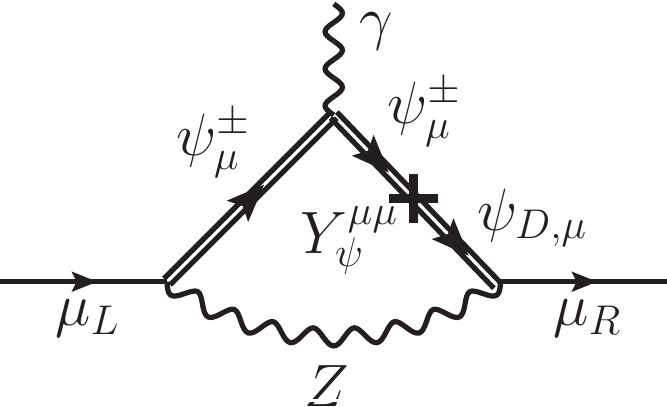}
\hspace*{0.2in}
\includegraphics[scale=0.65]{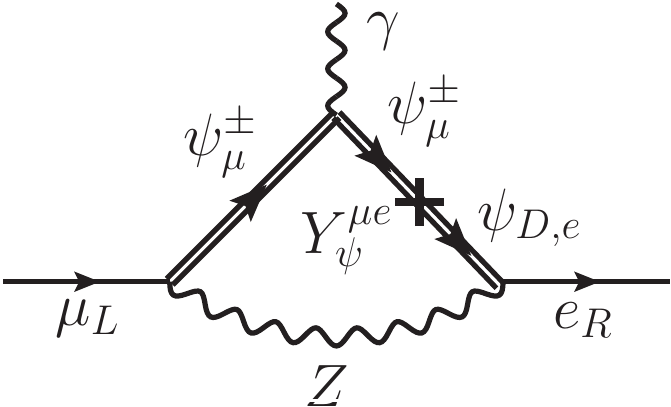}
\hspace*{0.2in}
\includegraphics[scale=0.65]{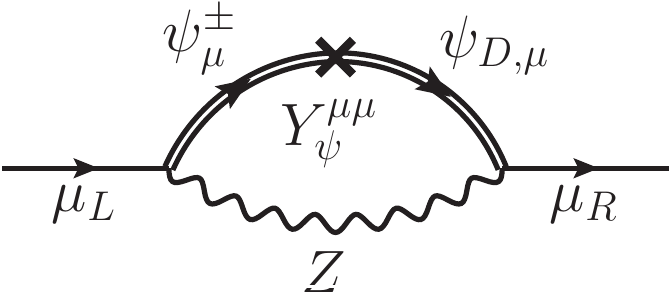}
\caption{Contributions of new heavy leptons to $a_{\mu}$, $\mathcal{B}(\mu\rightarrow e\gamma)$, and $m_{\mu}$ (from left to right). Shown are representative diagrams for the case of weak charged-singlet ($\psi^{\pm}$) and doublet ($\psi_D$) leptons. The indices $e$ and $\mu$ denote positions $1$ and $2$ in flavor space, respectively.}
\label{fig:amu-meg-mmu}
\end{figure}

The lepton sector of the SM has an accidental approximate flavor symmetry $\mathcal{G}_F = SU(3)_L\times SU(3)_e$, under which weak doublet and charged singlet leptons transform as $(3,1)$ and $(1,3)$ representations, respectively. The flavor symmetry is broken only by the charged-lepton and neutrino Yukawa couplings ${\bf Y_{\ell}}$ and ${\bf Y_{\nu}}$,\footnote{Here ${\bf Y_{\nu}}$ generically refers to any fermion-scalar interaction responsible for neutrino mass generation, which may involve new scalars (as in type-II seesaw) or fermions (as in type-III seesaw).} a pattern referred to as minimal flavor violation. 

The presence of new vector leptons generally introduces new sources of lepton flavor violation through their mass term ${\bf M_{\psi}}$ and Yukawa coupling ${\bf Y_{\psi}}$ to SM leptons or other vector leptons. We extend the principle of MFV to vector leptons by making the following demands. Vector leptons transform under $\mathcal{G}_F$ as either $(3,1)$ or $(1,3)$ representations, which implies three flavor copies of each new vector lepton. Furthermore, ${\bf M_{\psi}}$ and ${\bf Y_{\psi}}$ must transform under $\mathcal{G}_F$ as appropriate combinations of ${\bf Y_{\ell}}=(3,\overline{3})$ and ${\bf Y_{\nu}}=(3,1)$. This principle applies accordingly to new vector bosons with gauge couplings ${\bf G}_V$ or scalars with couplings ${\bf G_{\phi}}$. In the eigenbasis of weak interactions, the masses and couplings of new particles thus respect the following pattern in flavor space,
\begin{align}
{\bf M_{\psi}} &=  m_{\psi} ({\bf 1} + c_M\,{\bf \Delta}'_{\psi})\,, & {\bf \ \ Y_{\psi}} &= y_{\psi}{\bf Y_{\ell}}(1 + c_{\psi}\,{\bf \Delta}_{\psi})\quad \text{or}\quad y_{\psi}({\bf 1} + c'_{\psi}\,{\bf \Delta}'_{\psi})\,,\\\nonumber
{\bf G}_V &=  g_V({\bf 1} + c_V\, {\bf \Delta}_V)\,, & {\bf \ \ G_{\phi}} &= g_{\phi}{\bf Y_{\ell}}(1 + c_{\phi}\,{\bf \Delta}_{\phi})\quad \text{or}\quad g_{\phi}({\bf 1} + c'_{\phi}\,{\bf \Delta}'_{\phi})\,,
\end{align}
where $y_{\psi}$, $g_V$, $g_{\phi}$, $c_i$ and $c'_i$ are arbitrary coefficients of $\mathcal{O}(1)$ and $m_{\psi}$ sets the scale for the masses of vector leptons. For our purposes, $c_M{\bf \Delta}'_{\psi}$ and $c_V{\bf \Delta}_{V}$ can be neglected, yielding flavor-universal masses ${\bf M_{\psi}} =  m_{\psi}\times {\bf 1}$ and gauge couplings ${\bf G}_V = g_V \times {\bf 1}$. Flavor violation is potentially induced by the matrices ${\bf \Delta}_i$, which are combinations of $\bf Y_{\ell}$ and $\bf Y_{\nu}$ of $\mathcal{O}({\bf Y}_{\ell,\nu}^2)$ and higher. The exact form of ${\bf \Delta}_i$, as well as the transformation properties of ${\bf Y_{\psi}}$ and ${\bf G_{\phi}}$ under the flavor group, depend on the representation of the (vector) leptons. In particular, the magnitude of the mixing between new vector leptons is determined by $\bf Y_{\psi}\sim \bf Y_{\ell}$ ($\bf Y_{\psi}\sim \bf 1$), if they are in different (in the same) representations of $\mathcal{G}_F$. The consequences on effects in $a_{\mu}$ will be discussed in Section~\ref{sec:twomixed}.

Under these conditions, contributions to $\mathcal{B}(\mu\rightarrow e\gamma)$ from vector leptons are suppressed by neutrino mass splittings (encoded in $\Delta$) as in the SM, but effects in flavor-conserving observables such as $a_{\mu}$ can be sizeable. In Figure~\ref{fig:amu-meg-mmu}, we illustrate contributions of vector leptons to $a_{\mu}$ (left) and $\mathcal{B}(\mu\rightarrow e\gamma)$ (center) for the case of a weak singlet $\psi^{\pm}=(3,1)$ and a doublet $\psi_D=(1,3)$. In the mass eigenbasis of the charged leptons, the Yukawa couplings are given by $Y_{\psi}^{\mu\mu} = y_{\psi} y_{\mu}$ and $Y_{\psi}^{\mu e} = \Delta_{\mu e} y_{\mu}$, where $\Delta_{\mu e}$ is proportional to the neutrino mass splittings.

Minimal flavor violation also protects the SM lepton masses from overly large quantum corrections induced by vector leptons. In general, the Yukawa mixing ${\bf Y_{\psi}}$ between vector leptons in different flavor representations induces potentially large contributions to the lepton masses, ${\bf M_{\ell}} = ({\bf Y_{\ell}} + {\bf Y_{\psi}} L)\, v/\sqrt{2}$, where $L$ is a loop factor of $\mathcal{O}(1/(4\pi))$.
These effects are illustrated in Figure~\ref{fig:amu-meg-mmu}, right. Within the framework of MFV, mass corrections are proportional to the lepton Yukawa coupling, yielding
\begin{eqnarray}
{\bf M_{\ell}} = {\bf Y_{\ell}} (1 + y_{\psi}L)\, v/\sqrt{2}\,.
\end{eqnarray}
Perturbativity imposes an upper bound of $|y_{\psi}|\lesim \sqrt{4\pi}/y_{\tau}\approx 3.5 \times 10^2$. For effects in the muon sector, the relevant Yukawa coupling is thus confined to $|Y_{\psi}^{\mu\mu}|=|y_{\psi}|y_{\mu} \lesim 0.2$. If vector leptons are in the same flavor representation, their mixing $Y_{\psi}$ is unconstrained by MFV. In this case, the muon mass is protected by the suppressed mixing of vector leptons with SM leptons, which will be discussed in Section~\ref{sec:twomixed}.


\subsection{Calculational techniques and tools}
The calculation of our new electroweak contributions to $a_{\mu}$ is performed in a semi-automated way. We generate the one-loop amplitudes for the process $\mu\rightarrow\mu\gamma$ in the unitary gauge using the {\sc FeynArts} package \cite{Hahn:2000kx}, supplemented by the Feynman rules for the new particles. The calculation of the contributions to $a_{\mu}$ is greatly simplified by applying a projection technique that singles out the magnetic form factor \cite{Czarnecki:1996if}. Subsequently, amplitudes are evaluated for zero momentum transfer and expanded up to leading order, ${\cal O}(m_\mu^2)$, in the small muon mass (or, equivalently, the muon Yukawa coupling). This procedure, as well as the reduction of the loop integrals, has been performed with two independent computer programs, one of which is based on {\sc FeynCalc} \cite{Mertig:1990an}, while the other is a private code. We thereby have obtained a cross check of all analytic results.

By assuming that the correction to $a_\mu$ in a given new-physics scenario can explain the observed discrepancy in \eqref{d}, we obtain constraints on the parameter space of particle masses and couplings. In some cases, the correction turns out to have the wrong sign or is generically too small. As described in the next sections, we still find a number of scenarios that provide a successful explanation for the discrepancy. We then analyze the production mechanism and typical decay signatures of the new particles at the LHC. For this purpose, we do not assume any additional particle content and couplings besides those appearing in the $a_\mu$ loop corrections or required by gauge invariance. Cross sections and event rates are computed at the parton level using the program {\sc CalcHEP} \cite{calchep}. We then aim at setting bounds on the allowed parameter space from LHC data by recasting existing BSM searches of the ATLAS and CMS collaborations.


\section{One new field}\label{sec:onefield}

This section discusses scenarios where a single new field (that couples to
muons) at a time is added to the SM. For all fields listed in Table~\ref{tab:notation}, we analyze their contributions to $a_{\mu}$ and potential constraints from LEP observables. Subsequently, we identify the parameter space that can explain the discrepancy $\Delta a_{\mu}$. Analytic results for the contributions to $a_\mu$ are summarized in Table~\ref{tab:onep} in Appendix~\ref{sec:olf}.

\begin{figure}
\centering
\includegraphics[scale=0.8]{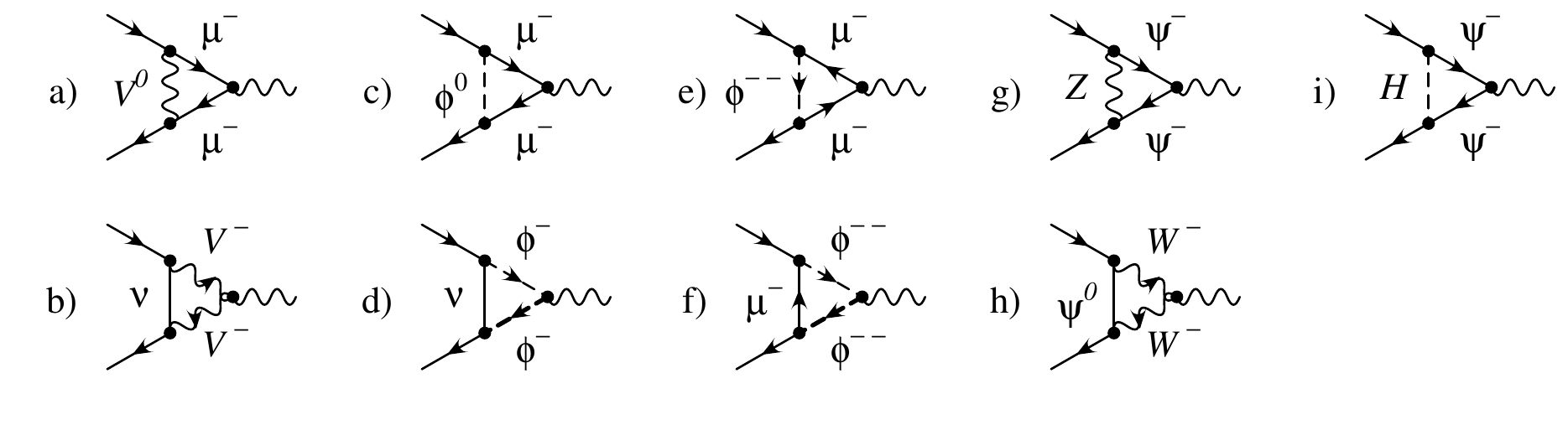}
\caption{Electroweak contributions to $a_{\mu}$ with one new particle in the vertex loop.}
\label{fig:onep}
\end{figure}

\paragraph{Neutral vector boson ($V^0$):}
A massive neutral vector boson with the effective couplings to leptons of the form
\begin{equation}
{\cal L} \supset g_L \overline{\ell_L} \gamma^\mu \ell_L V^0_\mu
 + g_R \overline{\ell_R} \gamma^\mu \ell_R V^0_\mu
\label{v0}
\end{equation}
can contribute to $a_\mu$ through the diagram in Figure~\ref{fig:onep}~(a).
The correction $\delta a_\mu$ becomes maximal for $g_L=g_R\equiv g_V$, for fixed
$\sqrt{g_L^2+g_R^2}$, in which case
the discrepancy in \eqref{d} can be explained for 
\begin{equation}
0.0047\gev^{-1} < g_V/M_V < 0.0062\gev^{-1}
\end{equation}
at the one-sigma level. As long as $V^0$ does not mix with the $Z$ boson, constraints from $Z$-pole precision observables at LEP can be evaded. However, assuming MFV, the interaction
\eqref{v0} generates $ee\mu\mu$ and $eeee$ contact interactions, which have been
strongly constrained by the LEP experiments at CM energies of $\sqrt{s}\approx 130 - 200\gev$. For $M_V > \sqrt{s}$, the limit from \cite{lep4} on the
scale $\Lambda$ of the $ee\mu\mu$ operator can be translated into the $95\%$ C.L.\ upper bound
\begin{eqnarray}\label{eq:gvci}
g_V/M_V = \sqrt{4\pi}/\Lambda < 0.00022\gev^{-1}\,,\qquad M_V > 200\gev\,.
\end{eqnarray}
For $M_V < \sqrt{s}$, neutral vector bosons can be resonantly produced via $e^+e^- \rightarrow V^0\gamma \rightarrow \ell^+\ell^-\gamma$, where $\gamma$ is a soft or hard photon. The cross section for the production of a narrow resonance $R$ with a total width $\Gamma_R$ is proportional to 
\begin{eqnarray}\label{eq:bwres}
\sigma(e^+e^- \rightarrow R\gamma \rightarrow \ell^+\ell^-\gamma) \propto \frac{2j+1}{\Gamma_R}\Gamma(R\rightarrow e^+ e^-)\Gamma(R\rightarrow\ell^+\ell^-)\,,
\end{eqnarray}
with $j=1 (0)$ for a vector (scalar) resonance. The partial decay widths of vectors and scalars into leptons are given by $\Gamma(V\rightarrow \ell^+\ell^-) = g_{\ell}^2M_V/(24\pi)$ and $\Gamma(\phi\rightarrow \ell^+\ell^-) = g_{\ell}^2M_{\phi}/(16\pi)$, respectively. At LEP, resonance searches for scalar neutralinos with $R$-parity violating coup\-lings $\lambda$ have been performed at CM energies in the range of $\sqrt{s} = 130\dots 189\gev$ \cite{Abbiendi:1999wm}. For a decay width $\Gamma_{\tilde{\nu}} \le 1\gev$, the couplings to leptons are constrained to $\lambda < 0.02\dots0.08$ at the $95\%$ C.L., depending on the neutralino mass $M_{\tilde{\nu}}$, in the mass range $100\gev < M_{\tilde{\nu}} < 200\gev$. Interpreting the bounds on $\lambda$ for vector bosons and fixing the total decay width to $\Gamma_V = 1\gev$,\footnote{For larger decay widths, the bound on $g_V$ is mildly relaxed. For instance, for $\Gamma_V = 10\gev$, resonance searches yield $g_V < 0.14$, which is still below the range required to explain $\Delta a_{\mu}$.} yields the conservative $95\%$ C.L.\ upper bound 
\begin{eqnarray}\label{eq:gvres}
g_V/M_V < 0.08 \times \sqrt[4]{3/4}/M_V \lesim 0.00075\gev^{-1}\,,\qquad 100\gev < M_V < 200\gev\,.
\end{eqnarray}%
The bounds from contact interactions \eqref{eq:gvci} and resonance searches \eqref{eq:gvres} at LEP therefore rule out sizeable contributions to $a_{\mu}$ from neutral vector bosons. For the same reasons, any SU(2) multiplet of vector bosons containing a neutral
vector field is excluded.

\begin{figure}
\includegraphics[scale=0.78]{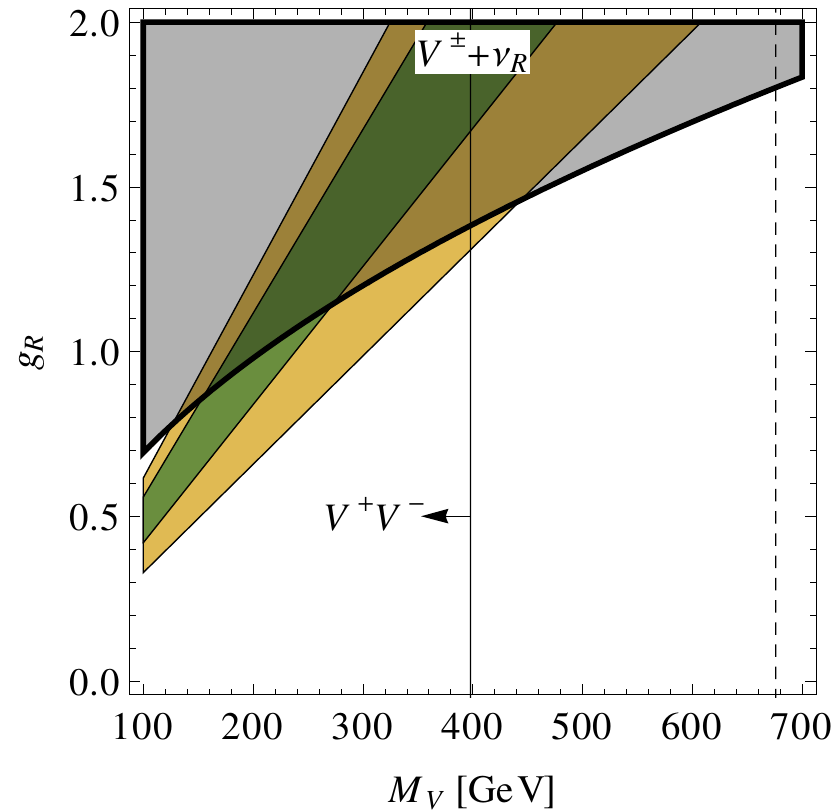}
\hfill
\begin{minipage}[b]{9cm}
\caption{Contribution to $a_{\mu}$ from a charged vector boson $V^{\pm}$ and a light right-handed neutrino $\nu_R$ in the vertex loop. The parameter space to explain $\Delta a_{\mu}$ at the $1\sigma$ ($2\sigma$) level is displayed in green (yellow). The $95\%$ C.L. region excluded by $ee\ell\ell$ contact interaction searches at LEP is shaded gray. Lower mass bounds at $95\%$ C.L. from direct searches at the 8-TeV LHC and projections for $14\tev$ (see Section~\ref{sec:lhc}) are displayed as plain and dashed black lines, respectively.\newline\newline}
\label{fig:chargedvector}
\end{minipage}
\end{figure}

\paragraph{Charged vector boson ($V^\pm$):} A charged vector boson can
contribute to $a_\mu$ through the diagram in Figure~\ref{fig:onep}~(b). Since limits from electroweak precision
tests are stronger for a coupling of $V^\pm$ to left-handed SM fermions than to right-handed fermions (due
to interference with the $W$ boson), the latter case
is considered here,
\begin{equation}
{\cal L} \supset g_R \overline{\ell_R} \gamma^\mu \nu_R V^-_\mu + \text{h.c.}
\label{vc}
\end{equation}
We do not speculate about the nature of the right-handed neutrino and assume it to be light ($M_{\nu_R} \ll M_{\text{EW}}$),\footnote{The case of weak-singlet neutrinos $\nu_R=\psi^0$ with $M_{\psi^0} \approx M_{\text{EW}}$ will be discussed in Section~\ref{sec:twofields}.} but still heavy enough ($M_{\nu_R}\gesim 100\mev$) to evade potential bounds from muon decay and astrophysics. In this range, the contribution to $a_{\mu}$ is well approximated by $M_{\nu_R}=0$, and the discrepancy \eqref{d} can a priori be explained at the one-sigma level for 
\begin{equation}
0.0042\gev^{-1} < g_R/M_V < 0.0056\gev^{-1}\,.
\end{equation}
The corresponding parameter space is displayed in Figure~\ref{fig:chargedvector}. Constraints on $V^{\pm}$ contributions to $a_{\mu}$ are derived from bounds on four-lepton contact interactions at LEP. The leading effect occurs at the one-loop level through the box diagram in Figure~\ref{fig:box}~(c). This effect yields the $95\%$ C.L. bound $g_R^2/M_V<0.0048\gev^{-1}$, which excludes the parts of the parameter space corresponding to the gray region in Figure~\ref{fig:chargedvector}. One-loop four-lepton interactions will be discussed in detail in Section~\ref{sec:twofields}, where they play a crucial role in constraining couplings of SM leptons to two new fields with different spins.

\paragraph{Scalar doublet ($\phi_D$):}
For a scalar doublet, one can write down lepton couplings similar to the Yukawa couplings of
the SM Higgs boson,
\begin{equation}\label{eq:scalardoublet}
{\cal L} \supset -Y\overline{L_L}\phi_D\ell_R + \text{h.c.}, \qquad
\phi_D = \begin{pmatrix}
 \phi_D^+ \\ \phi_D^0
\end{pmatrix},
\end{equation}
where $L_L$ is the left-handed SM lepton doublet and
$\phi_D^{+,0}$ are the charged and neutral (complex) components of $\phi_D$,
respectively. It is assumed that $\phi_D$ does not have a vev that would contribute to fermion masses.
The scalar doublet can contribute to $a_\mu$ through the diagrams
Figure~\ref{fig:onep}~(c,d). It turns out that $\phi_D$ can successfully accommodate $\Delta a_{\mu}$ for
\begin{equation}\label{eq:gsamu}
0.0076\gev^{-1} < Y/M_\phi < 0.0102\gev^{-1}
\end{equation}
at the one-sigma level. As for a neutral vector boson, the exchange of a neutral scalar in $e^+e^-$ collisions generates four-lepton contact interactions for $M_{\phi} > \sqrt{s}$. Direct constraints on scalar four-fermion contact interactions from LEP do not exist. Still, the bounds on $eeee$ vector interactions can be interpreted as bounds on scalar interactions by using the Fierz identity
\begin{eqnarray}
(\overline{e_R}e_L)(\overline{e_L}e_R) + (\overline{e_L}e_R)(\overline{e_R}e_L) = \frac{1}{2}\big[(\overline{e_L}\gamma_{\mu}e_L)(\overline{e_R}\gamma_{\mu}e_R) + (\overline{e_R}\gamma_{\mu}e_R)(\overline{e_L}\gamma_{\mu}e_L)\big]\,.
\end{eqnarray}
The limits from \cite{lep4} on the scale $\Lambda_{\text{LR}}$ of the LR (and RL) four-electron vector operator thus translate into the $95\%$ C.L. limit
\begin{eqnarray}\label{eq:gsci}
Y/M_{\phi} = \sqrt{2\pi}/\Lambda_{\text{LR}} < 0.00025\gev^{-1}\,,\qquad M_{\phi} > 200\gev\,.
\end{eqnarray}

For $M_{\phi} < \sqrt{s}$, the LEP searches for neutralino resonances discussed around \eqref{eq:bwres} apply directly to neutral scalars. They lead to the $95\%$ C.L. upper bound
\begin{eqnarray}\label{eq:gsres}
Y/M_{\phi} < 0.08/M_{\phi} \lesim 0.0008\gev^{-1}\,,\qquad 100\gev < M_{\phi} < 200\gev\,.
\end{eqnarray}
By comparing the bounds from \eqref{eq:gsres} and \eqref{eq:gsci} with \eqref{eq:gsamu}, it is evident that a scalar doublet as an explanation of $\Delta a_{\mu}$ is ruled out by LEP searches for neutral scalars.

\paragraph{Scalar triplet ($\phi_T$):}
A scalar triplet $\phi_T$ with hypercharge $-1$ can couple to muons through the
interaction
\begin{equation}\label{eq:scalartriplet}
{\cal L} \supset -\frac{Y}{2}\overline{L_L}\phi_T i\sigma_2L_L^c + \text{h.c.},
\qquad
\phi_T = \begin{pmatrix}
 \phi_T^-/\sqrt{2} & \phi_T^0 \\ \phi_T^{--} & -\phi_T^-/\sqrt{2}
\end{pmatrix},
\end{equation}
where $\sigma_2$ is the second Pauli matrix. The correction $\delta a_\mu$, corresponding to the diagrams in Figure~\ref{fig:onep}~(c-f), is always negative and thus cannot
explain the observed discrepancy $\Delta a_{\mu}$.

\paragraph{Vector-like fermions ($\psi^0$, $\psi^\pm$, $\psi_D$, $\psi_A$, $\psi_T$):} 
New fermions with vector-like mass terms can couple to the SM leptons through Yukawa
couplings involving the SM Higgs doublet $H$. We consider the following cases:
\begin{itemize}
\item A neutral SU(2) singlet $\psi^0$;
\item A charged SU(2) singlet $\psi^\pm$;
\item An SU(2) doublet $\psi_D$ with the same quantum numbers as the left-handed SM lepton
doublet;
\item An SU(2) triplet $\psi_A$ with hypercharge 0 ($i.\,e.$ in the adjoint
representation) and Majorana mass term;
\item An SU(2) triplet $\psi_T$ with hypercharge $-1$.
\end{itemize}
The relevant Yukawa couplings for these five cases are given by
\begin{align}
&{\cal L} \supset -Y\overline{L_L}\tilde{H}\psi^0_R + \text{h.c.}, \label{fn} \\
&{\cal L} \supset -Y\overline{L_L}H\psi^-_R + \text{h.c.}, \label{fc} \\
&{\cal L} \supset -Y\overline{\psi_{D,L}}H\ell_R + \text{h.c.}, &&
\psi_D = \begin{pmatrix}
 \psi_D^0 \\ \psi_D^-
\end{pmatrix}, \label{fd} \\
&{\cal L} \supset -Y \tilde{H}^\dagger \overline{\psi_{A,R}} L_L + \text{h.c.},
&&
\psi_A = \begin{pmatrix}
 \psi_A^0/\sqrt{2} & \psi_A^+ \\ \psi_A^{-} & -\psi_A^0/\sqrt{2}
\end{pmatrix}, \qquad\qquad \label{fa} \\
&{\cal L} \supset -Y H^\dagger \overline{\psi_{T,R}} L_L + \text{h.c.}, &&
\psi_T = \begin{pmatrix}
\psi_T^-/\sqrt{2} & \psi_T^{0} \\ \psi_T^{--} & -\psi_T^{-}/\sqrt{2}
\end{pmatrix}, \label{ft}
\end{align}
where $\tilde{H} = i\sigma_2 H^*$. After
electroweak symmetry breaking, when $H$ acquires a vev $\langle H \rangle = (0,v/\sqrt{2})^\top$, these interactions lead to
mixing between the vector-like fermions and the SM charged leptons or neutrinos, which can be expressed in terms of the
mixing parameter $\epsilon = Yv/M_\psi$. The mixing affects the electroweak couplings of SM leptons by corrections of ${\cal O}(\epsilon^2)$ and induces new gauge and Yukawa interactions of a vector lepton with a SM boson and a SM lepton of ${\cal O}(\epsilon)$. The former effect modifies the size of the SM electroweak contributions to $a_\mu$, whereas the new couplings lead to additional contributions to $a_{\mu}$ from the diagrams in Figure~\ref{fig:onep}~(g),~(h) and/or (i). The corrections to $a_{\mu}$ are of ${\cal O}(\epsilon^2)$ in either case. Details on vector lepton mixing and the resulting electroweak couplings in the context of $a_{\mu}$ can be found, for instance, in \cite{Kannike:2011ng,Dermisek:2013gta}.

The analytic results for effects of mixing vector leptons on $a_{\mu}$ are listed in Table~\ref{tab:onep} in Appendix~\ref{sec:olf}. For the neutral singlet $\psi^0$ and the triplets $\psi_A$, $\psi_T$, the correction $\delta a_\mu$ is negative. For the charged singlet $\psi^\pm$, $\delta a_{\mu}$ is positive for $M_\psi \gesim 250\gev$, but too small to explain the observed discrepancy with perturbative couplings $|Y| < \sqrt{4\pi}$. The contribution of the doublet $\psi_D$ can a priori accommodate $\Delta a_{\mu}$ for strong mixing $|\epsilon| \gesim 1.2$ and perturbative couplings in the mass range $100\gev < M_\psi < 500\gev$.
However, the mixing between SM leptons and heavy vector leptons is strongly constrained by $Z$-pole precision measurements at LEP. Assuming flavor-universal couplings, a global fit to LEP data leads to the bound $|\epsilon| \lesim 0.03$ for mixing with a vector lepton doublet \cite{delAguila:2008pw}, clearly ruling out any significant contribution to $a_{\mu}$.


\section{Two new mixed fermion fields}\label{sec:twomixed}

In the previous Section~\ref{sec:onefield}, we have seen that effects on $a_{\mu}$ from a single species of vector-like fermions are either negative or too small to explain the discrepancy $\Delta a_{\mu}$ in (\ref{d}).
However, larger corrections may in principle be obtained from the simultaneous presence of two types of vector leptons that mix with each other \cite{Kannike:2011ng,Dermisek:2013gta}. Possible combinations in accord with weak quantum numbers are a weak doublet $\psi_D$ with either a neutral singlet $\psi^0$, a charged singlet $\psi^{\pm}$, 
a weak adjoint triplet $\psi_A$, or a triplet $\psi_T$ with hypercharge $-1$.

In addition to the mixing with SM fermions in (\ref{fn})--(\ref{ft}), vector leptons with dif\-fe\-rent weak quantum numbers mix through Yukawa couplings to the SM Higgs boson. The Lagrangian describing the mixing of a doublet with a singlet or a triplet reads
\begin{align}\label{eq:DSmixing}
{\cal L}^{\text{mix}}_{DS} &= - Y_{DS}\overline{\psi_{D,L}}H\psi^-_R - Y_{SD}\overline{\psi^-_L}H^{\dagger}\psi_{D,R} + \text{h.c.}\\
{\cal L}^{\text{mix}}_{DN} &= - Y_{DN}\,\overline{\psi_{D,L}}\widetilde{H}\psi^0_R - Y_{ND}\,\overline{\psi^0_L}\widetilde{H}^{\dagger}\psi_{D,R} + \text{h.c.}\\
{\cal L}^{\text{mix}}_{DA} &= - Y_{DA}\overline{\psi_{D,L}}\psi_{A,R}\widetilde{H} - Y_{AD}\widetilde{H}^{\dagger}\overline{\psi_{A,L}}\psi_{D,R} + \text{h.c.}\\
{\cal L}^{\text{mix}}_{DT} &= - Y_{DT}\,\overline{\psi_{D,L}}\psi_{T,R}H - Y_{TD}\,H^{\dagger}\overline{\psi_{T,L}}\psi_{D,R} + \text{h.c.} \label{eq:DTmixing}
\end{align}
The required chirality flip in $a_{\mu}$ can thus proceed through the mixing between heavy leptons $(\sim Y_{12}v)$ rather than muons $(\sim y_{\mu}v)$, as illustrated in Figure~\ref{fig:amu-meg-mmu}, left.\footnote{$Y_{12}$ stands for either of the Yukawa couplings $Y_{SD}$, $Y_{DS}$, etc. inducing vector lepton mixing.} Contributions to $a_{\mu}$ from mixed vector leptons are thus enhanced by a factor of $Y_{12}/y_{\mu}$ with respect to contributions from single vector leptons. The complete analytic results for $a_{\mu}$ in the scenarios $\psi_D+\psi^{\pm}$, $\psi_D+\psi^0$, $\psi_D+\psi_A$, and $\psi_D+\psi_T$ are listed in Appendix~\ref{sec:olf} in (\ref{eq:amuWZH}) and (\ref{eq:amuWCC}); the corresponding couplings are defined in Tables~\ref{tab:mixedDS} and \ref{tab:mixedDA}. They are obtained by diagona\-li\-zing the mass matrices with mixing leptons $\ell$ and $\psi_1$ or $\psi_2$ to first order in the parameters $\epsilon_{i}=Y_{i}v/M_{i}$ (the mixing of SM leptons $\ell$ with vector leptons $\psi_{i}$) and $\omega_{12}=Y_{12}v/(M_1-M_2)$ (the mixing among vector leptons $\psi_1$ and $\psi_2$). We thereby retain the leading effects on $a_{\mu}$ up to $\mathcal{O}(\epsilon^2\omega)$ for moderate mixing $|\epsilon_{1,2}|,|\omega_{12}|\lesim 1$. The overall structure of $a_{\mu}$ can be expressed as the sum of contributions from single vector leptons and contributions from mixed vector leptons,
\begin{equation}\label{eq:amumix}
a_{\mu}(\psi_1,\psi_2) = m_{\mu}^2\epsilon_1^2\,F_1(M_1^2) + m_{\mu}^2\epsilon_2^2\,F_2(M_2^2) + m_{\mu}M_{1,2}\epsilon_1\epsilon_2\omega_{12}\,G(M_1^2,M_2^2)\,,
\end{equation}
where $F$ and $G$ are functions of the vector lepton masses $M_1$, $M_2$ and their couplings to SM bosons. Due to the enhancement of contributions with vector lepton mixing, the main effect on $a_{\mu}$ is to a good approximation given by the third term in (\ref{eq:amumix}). Without any further assumptions, the discrepancy $\Delta a_{\mu}$ can be accommodated for $M_{1,2} > 100\gev$ and couplings of $\mathcal{O}(0.1...1)$ in all scenarios.

The measurements of electroweak precision observables at LEP constrain the mixing with SM leptons to $|\epsilon_{S,D,T}| \lesim 0.03$ and $|\epsilon_{N,A}|\lesim 0.05$ for flavor-universal coup\-lings \cite{delAguila:2008pw}. In the framework of MFV, additional constraints on the couplings depend on the flavor representation (see Section~\ref{sec:amu-meg-mmu}). We consider two MFV scenarios, which result in the suppression of either the mixing with SM leptons $\epsilon_i$ or the mixing among vector leptons $Y_{12}$. Here we discuss them exemplarily for the case of vector singlet--doublet mixing.
\begin{enumerate}
\item Vector leptons are in the same representation as the SM leptons they mix with, i.e. $\psi_D = (1,3)$ and $\psi^{\pm} = (3,1)$.\footnote{In scenarios with vector lepton triplets, these transform in the same way as vector lepton singlets under the flavor group.} The couplings to SM leptons from (\ref{fc}) and (\ref{fd}) are thus flavor-conserving and the mixing parameter $\epsilon$ is unconstrained by MFV. The (flavor-breaking) mixing between $\psi_D$ and $\psi^{\pm}$ is proportional to the muon Yukawa coupling, $Y_{12} = y_{12}y_{\mu}$ ($12=SD,DS$). Including LEP constraints and requiring perturbativity, the couplings are limited to $|\epsilon_{S,D}| \lesim 0.03$ and $|Y_{12}| \lesim 0.2$.
\item Vector leptons are in the same representation, $\psi_D,\psi^{\pm} = (1,3)$, or $\psi_D,\psi^{\pm} = (3,1)$. In this case, only the coupling between $\psi^{\pm}$ ($\psi_D$) and SM leptons breaks the flavor symmetry, yielding the bound $|\epsilon_{S(D)}|M_{S(D)}/v = |Y_{S(D)}|y_{\mu}\lesim 0.2$. Since LEP limits are stronger than the MFV suppression in the mass range up to $M_{S,D}\sim 1.7\tev$, the couplings in either case are eventually limited to $|\epsilon_{S,D}|\lesim 0.03$. The mixing among vector fermions is unconstrained by the requirement of MFV, yielding $|Y_{12}|\lesim \sqrt{4\pi}$.
\end{enumerate}
In scenario 1, the maximal contributions to $a_{\mu}$ are of $\mathcal{O}(10^{-10})$, which is one order of magnitude too small to accommodate $\Delta a_{\mu}$ in (\ref{d}) within two sigma. In scenario 2, the discrepancy may a priori be explained by vector leptons around $M_{\rm EW}$ with sizeable mixing $Y_{12} \gesim 0.5$ in all four scenarios.

However, constraints on vector lepton mixing arise from the anomalous magnetic moment of the electron $a_e$. The discrepancy between the precise measurement and SM prediction has been found to be \cite{Aoyama:2012wj}
\begin{equation}\label{eq:ae}
\Delta a_{e} \equiv a_{e}^{\rm exp} - a_e^{\rm th} = (-1.06\pm 0.82)\times 10^{-12}\,.
\end{equation}
Within the framework of MFV, effects of mixing vector leptons on $a_{\mu}$ and $a_{e}$ are tightly related. The dominant contribution to $a_{\ell}$ in scenario 2 is proportional to $\delta a_{\ell}\sim m_{\ell}Y_1 Y_2 Y_{12}$, where either of the couplings to SM leptons is suppressed by the Yukawa coupling, $Y_i\sim y_{\ell}$, depending on the chosen flavor representations for $\psi_1$ and $\psi_2$. This leads to a quadratic scaling of vector-lepton effects on $a_{\ell}$ with the lepton masses, and in particular to $\delta a_{e}/\delta a_{\mu}=m_e^2/m_{\mu}^2$. However, LEP constraints on $Y_i$ are stronger than the MFV suppression $Y_i\sim y_{\ell}$ for masses below the TeV scale in the muon sector, but not in the electron sector. The maximal effect of mixing vector fermions in MFV therefore scales as
\begin{equation}\label{eq:aebound}
\frac{\delta a_e}{\delta a_{\mu}} = \frac{m_e^2}{m_{\mu}^2}\times \text{max}\Big[\frac{\sqrt{4\pi}y_{\mu}/y_{\tau}}{|\epsilon_i|^{\rm max}M_i/v},1\Big]\approx \frac{m_e^2}{m_{\mu}^2}\times \text{max}\Big[8.6\times \left(\frac{0.03\times 200\gev}{|\epsilon_i|^{\rm max}\times M_i}\right),1\Big].
\end{equation}

\begin{figure}
\centering
\begin{tabular}{cc}
\includegraphics[scale=0.78]{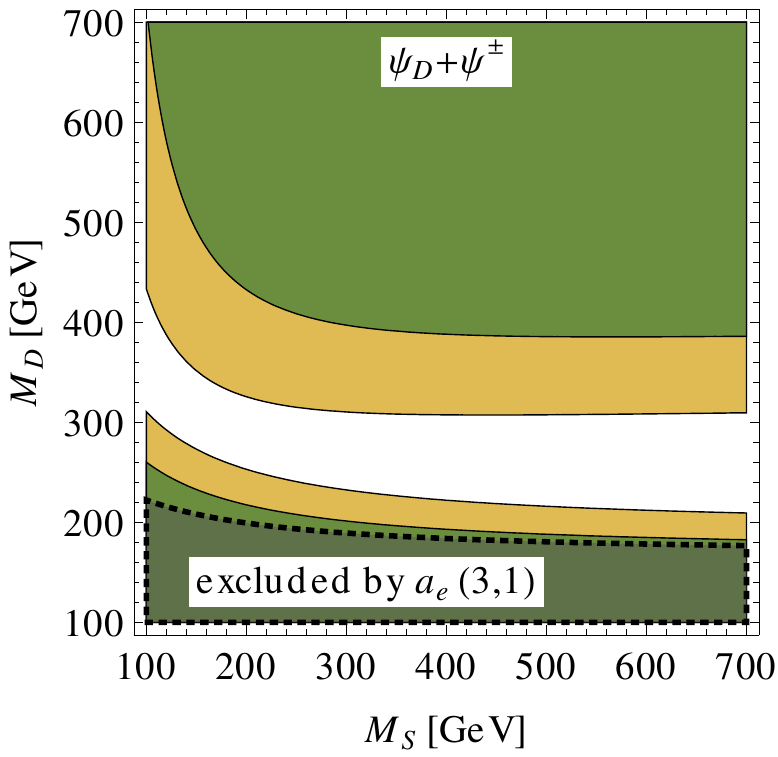}\hspace*{0.5cm} & \includegraphics[scale=0.78]{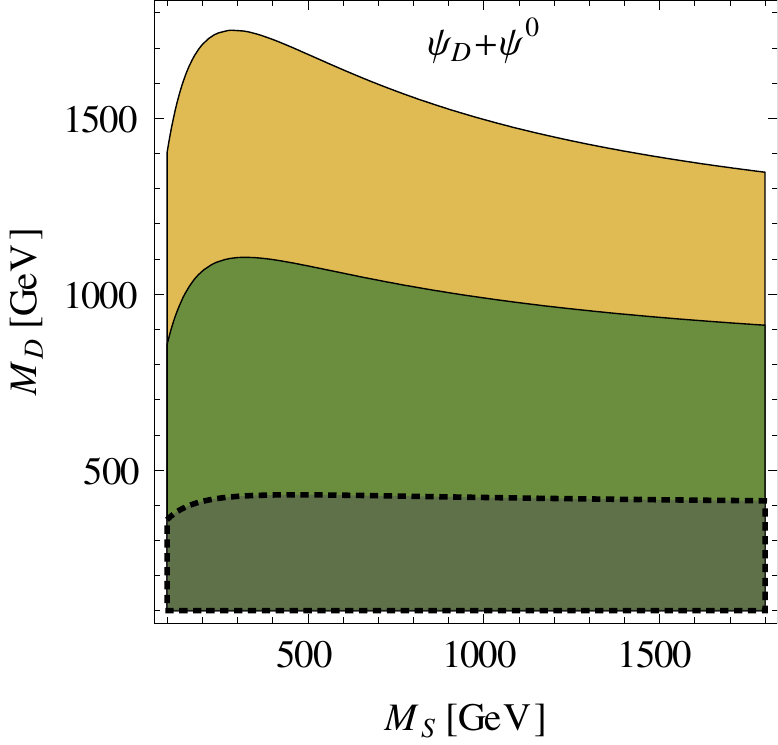}\\
\includegraphics[scale=0.78]{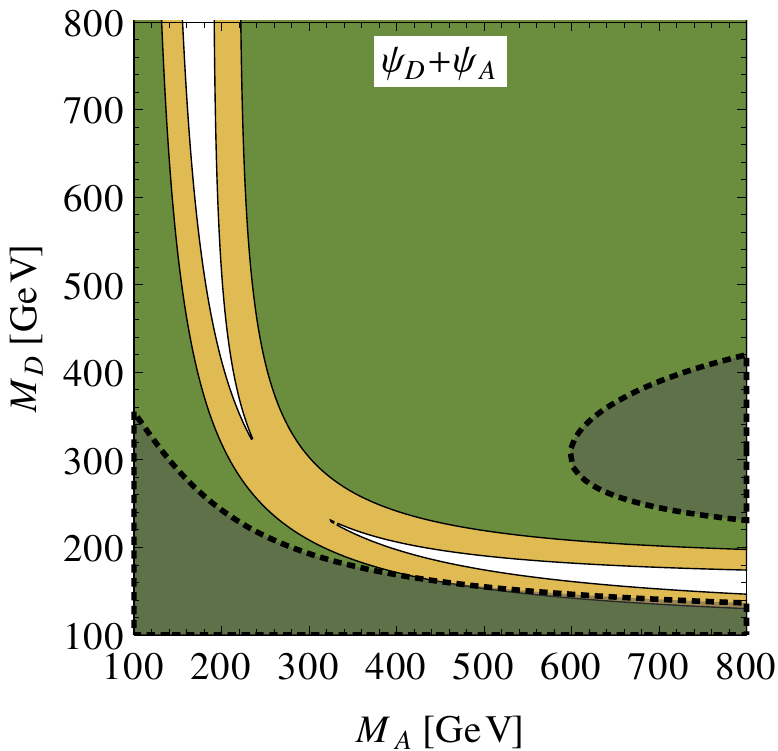}\hspace*{0.5cm} & \includegraphics[scale=0.78]{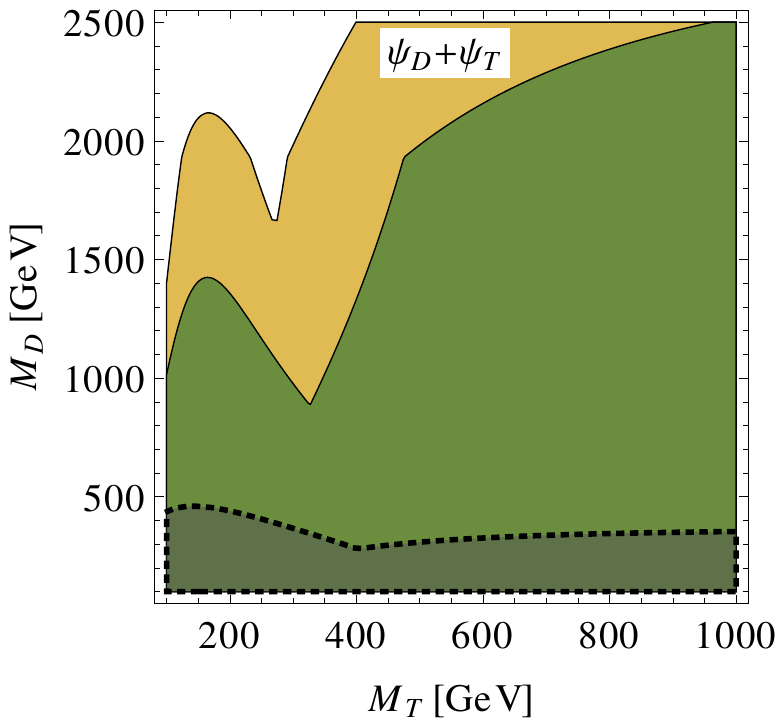}\\
\end{tabular}\vspace*{-0.3cm}
\caption{Contributions to $a_{\mu}$ from mixing vector leptons in the flavor scenario $(3,1)$. Regions that accommodate $\Delta a_{\mu}$ at the $1\sigma$ ($2\sigma$) level are displayed in green (yellow). Gray regions with dotted boundaries are excluded by $a_e$ at the $95\%$ C.L. in the scenario $(3,1)$.}
\label{fig:mix}
\end{figure}

In Figure~\ref{fig:mix}, we show the parameter space of mixing vector leptons that can explain $\Delta a_{\mu}$ at the $1\sigma$ (green) and $2\sigma$ (yellow) level in the flavor scenario 2.
 We choose the representation $(3,1)$, where both vector leptons are triplets under $SU(3)_L$, so that only the coupling $Y_D$ of $\psi_D$ to SM leptons is constrained by MFV. In the displayed parameter space, MFV bounds are visible only for the scenario $\psi_D+\psi_T$ as a kink in the one-sigma (two-sigma) region around $M_T\sim 500\gev$ ($270\gev$). Since the viable parameter space for $\Delta a_{\mu}$ below $M\lesim 2\tev$ is largely dominated by LEP bounds, similar bounds apply for the flavor representation $(1,3)$.

Bounds from $a_e$ are sensitive to the flavor representation, since MFV bounds compete with LEP bounds already in the low-mass region. For the case $(3,1)$, $a_e$ bounds exclude the gray area of low fermion-doublet masses. For the representation $(1,3)$, bounds from $a_e$ are of similar strength, but sensitive to the region with low singlet or triplet masses.

In the scenarios $\psi_D+\psi^{\pm}$, $\psi_D+\psi_A$ and $\psi_D+\psi_T$, the dominant contributions to $a_{\mu}$ decouple as {$Y_DY_iY_{12}/(M_DM_i)$, $i=S,A,T$,} for $M_{\psi} \gg M_{\rm EW}$. However, since LEP constraints on $Y_{1,2} = \epsilon_{1,2} M_{1,2}/v$ weaken as $M_{1,2}$ become large, the \emph{maximal} contribution to $a_{\mu}$ is asymptotically constant. In the scenario $\psi_D+\psi^0$, the dominant contribution due to vector fermion mixing decouples as $Y_NY_DY_{12}/M_D^2$ for $M_D \gg M_{\rm EW}$ and as $Y_NY_DY_{12}/M_N$ for $M_N \gg M_{\rm EW}$. The maximal $\delta a_{\mu}$ therefore decreases as $1/M_D$ for large doublet masses, but is constant in the limit of large singlet masses. In MFV, constraints on $Y_D\sim y_{\mu}$ exclude the parameter space with $M_D \gesim 4\tev$ in the scenario $\psi_D+\psi^{\pm}$, $M_D\gesim 7.6\tev$ in $\psi_D+\psi_A$, and $M_D\gesim 2.8\tev$ in $\psi_D+\psi_T$, and some of the parameter space with light singlets or triplets. Direct searches for vector fermions at the LHC are not able to probe the mass regime far above $M_{1,2}\sim 500-600\gev$ (see Section~\ref{sec:lhc}). Therefore an explanation of $\Delta a_{\mu}$ with mixing vector fermions, be it within or beyond MFV, cannot be excluded at the 14-TeV LHC.




\section{Two new fields with different spin}\label{sec:twofields}

Besides the case with two mixing fermions discussed in the previous section, two new fields with different spins can yield significant contributions to $a_{\mu}$. In this section, we discuss combinations of one vector fermion and one new scalar or vector boson. These two-field contributions to $a_{\mu}$ are dominant in scenarios where effects of a single new field are constrained by other observables or suppressed by symmetries. The corresponding Feynman diagrams are shown in Figure~\ref{fig:twop}; analytic expressions are given in Table~\ref{tab:twop} in Appendix~\ref{sec:olf}.

\begin{figure}
\centering
\includegraphics[scale=0.8]{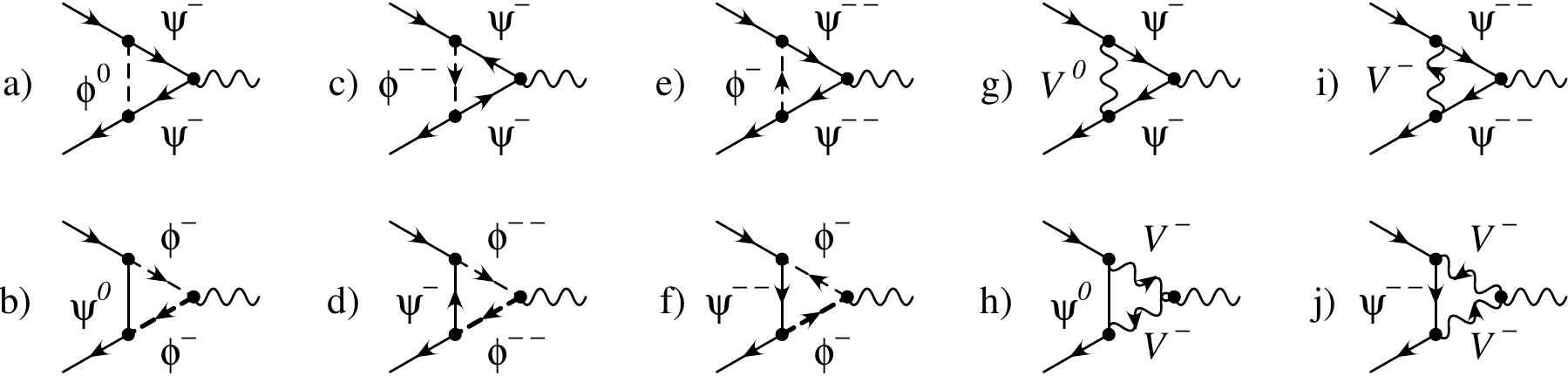}
\caption{Contributions to $a_{\mu}$ from two new particles with different spins in the vertex loop.}
\label{fig:twop}
\end{figure}

\begin{figure}
\centering
\begin{tabular}{ccccc}
\includegraphics[scale=0.62]{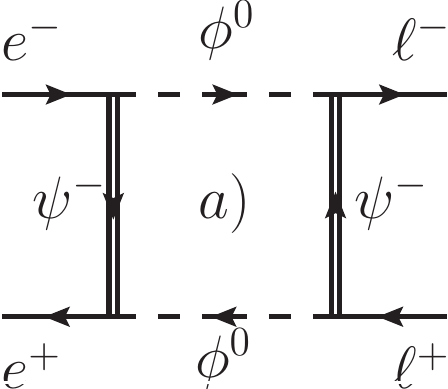}\hspace*{0.1cm} & \includegraphics[scale=0.62]{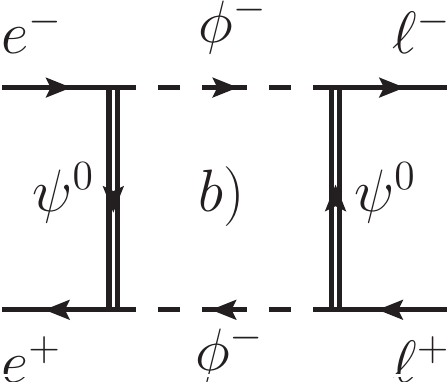}\hspace*{0.1cm} & \includegraphics[scale=0.62]{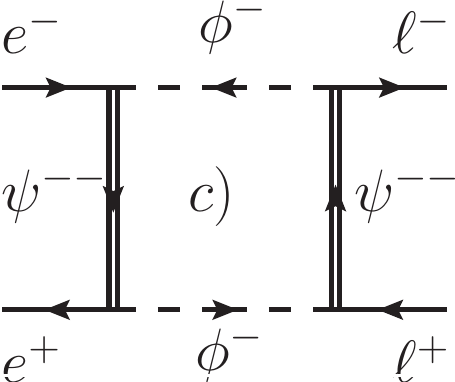}\hspace*{0.1cm} & \includegraphics[scale=0.62]{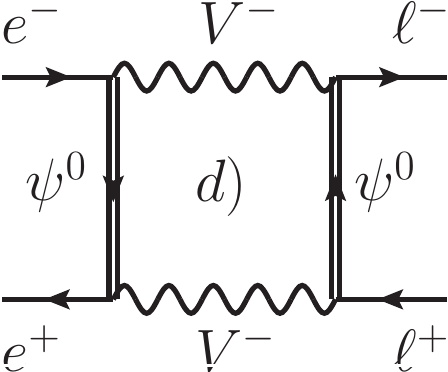}\hspace*{0.1cm} & \includegraphics[scale=0.62]{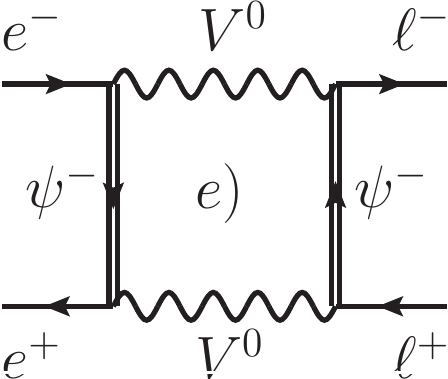}
\end{tabular}
\caption{One-loop contributions to effective four-lepton interactions $ee\ell\ell$.}
\label{fig:box}
\end{figure}

Constraints on the coupling of one SM lepton to a new vector fermion and a vector or scalar boson, $\overline{\ell}\,\psi\,V/\phi$, can be derived from $e^+e^-\rightarrow \ell^+\ell^-$ processes measured at LEP \cite{lep4}.\footnote{We restrict ourselves to leptons $\ell = \mu,\tau$ in the final state, which lead to stronger constraints on $LL$ and $RR$ interactions than $\ell = e$.} In the limit $M_{V,\phi,\psi}\gg \sqrt{s}$, new-physics effects in these processes can be described by effective four-lepton interactions 
\begin{align}
\mathcal{H}_{\rm eff} = \sum_{A,B=L,R}\mathcal{C}_{AB}\,\mathcal{O}_{AB}\,,\qquad \mathcal{O}_{AB} = (\overline{e}\gamma_{\mu}e_A)(\overline{\ell}\gamma^{\mu}\ell_B)\,,
\end{align}
where $\mathcal{O}_{AB}$ are local operators and $A,B=L,R$ indicate the chirality of the lepton fields. Two-particle couplings $\overline{\ell}\,\psi\,V/\phi$ generate four-lepton contact terms at the one-loop level through the box diagrams in Figure~\ref{fig:box}, with the corresponding Wilson coefficients $\mathcal{C}_{AB}\sim g^4/(16\pi^2 M_{V,\phi}^2)$. Due to the loop suppression, two-particle couplings are expected to be less constrained than the one-particle couplings discussed in Section~\ref{sec:onefield}, which induce four-lepton interactions at the tree level, yielding $\mathcal{C}_{AB}\sim g^2/M_{V,\phi}^2$.  As we will see, constraints from four-lepton interactions can still have a considerable impact on two-particle effects on $a_{\mu}$, in particular in scenarios where the coupling $g$ is sizeable. The four-lepton interaction terms for the two-particle combinations relevant in this section are listed in Table~\ref{tab:4lci} in Appendix~\ref{sec:4lci}. Let us discuss the different scenarios one by one.

\begin{figure}
\centering
\begin{tabular}{cc}
\includegraphics[scale=0.78]{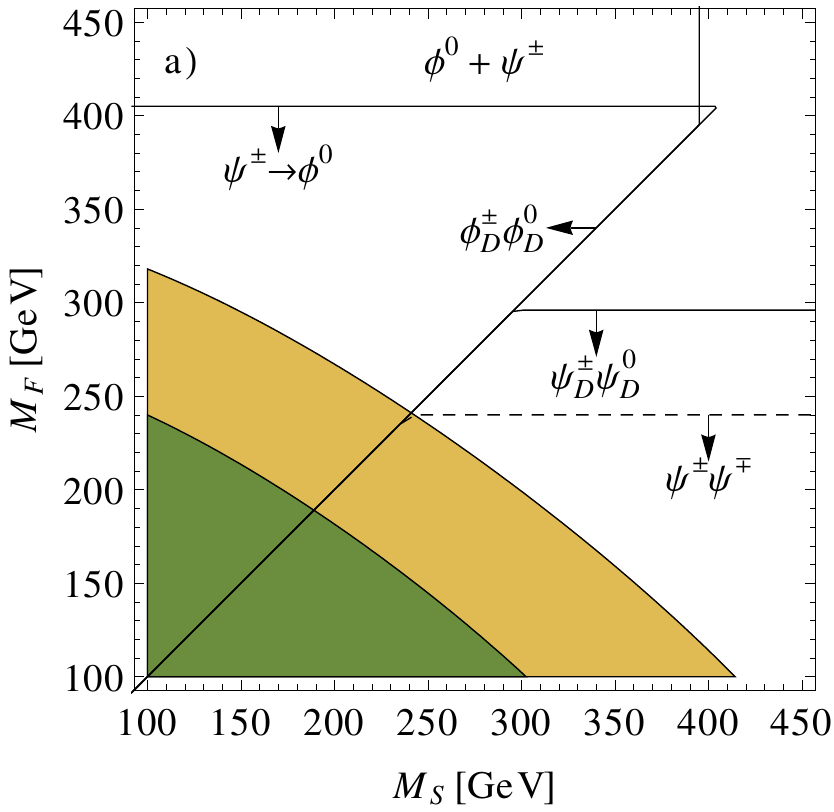}\hspace*{0.5cm} & \includegraphics[scale=0.78]{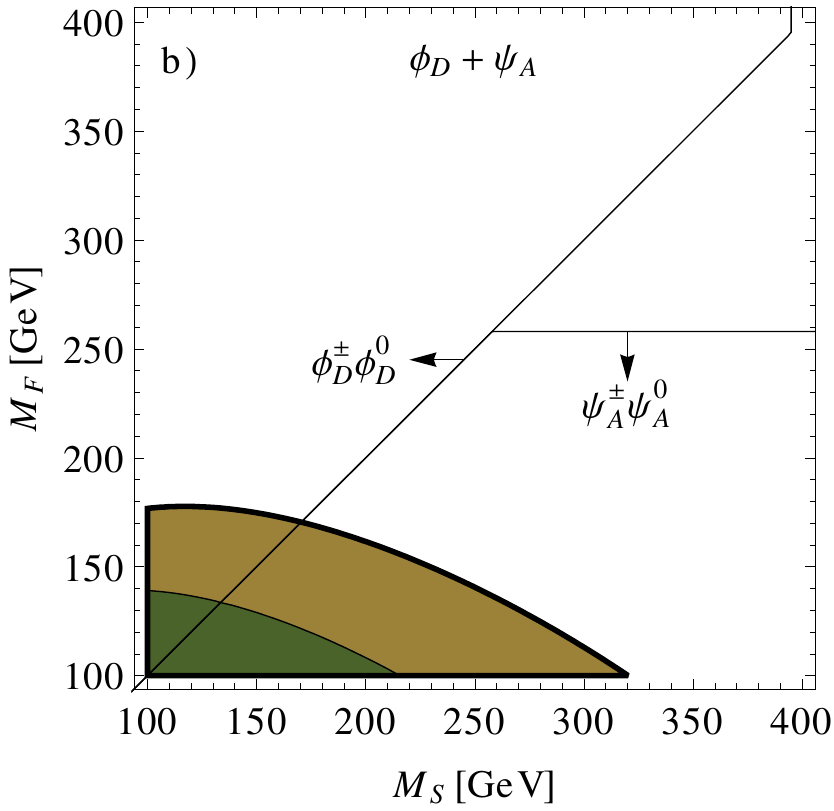}\\
\includegraphics[scale=0.78]{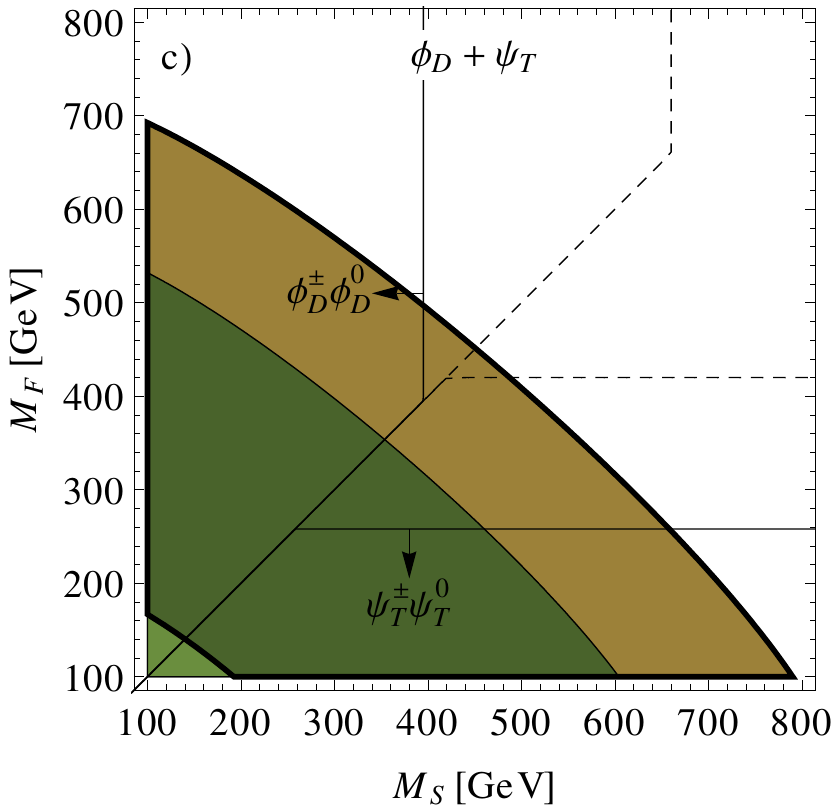}\hspace*{0.5cm} & \includegraphics[scale=0.78]{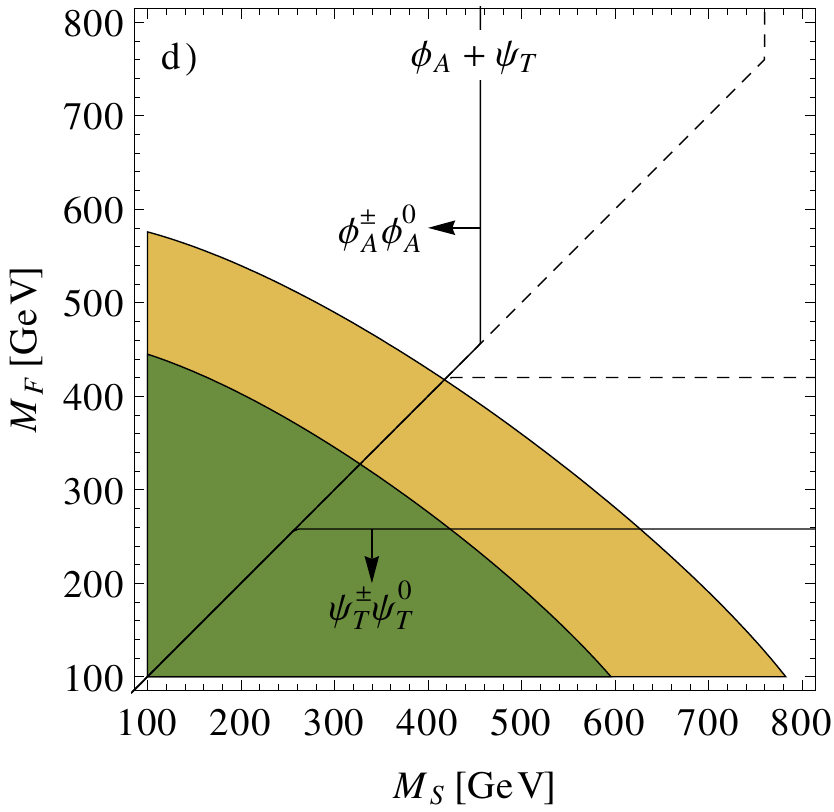}\\
\hspace*{-0.1cm}\includegraphics[scale=0.76]{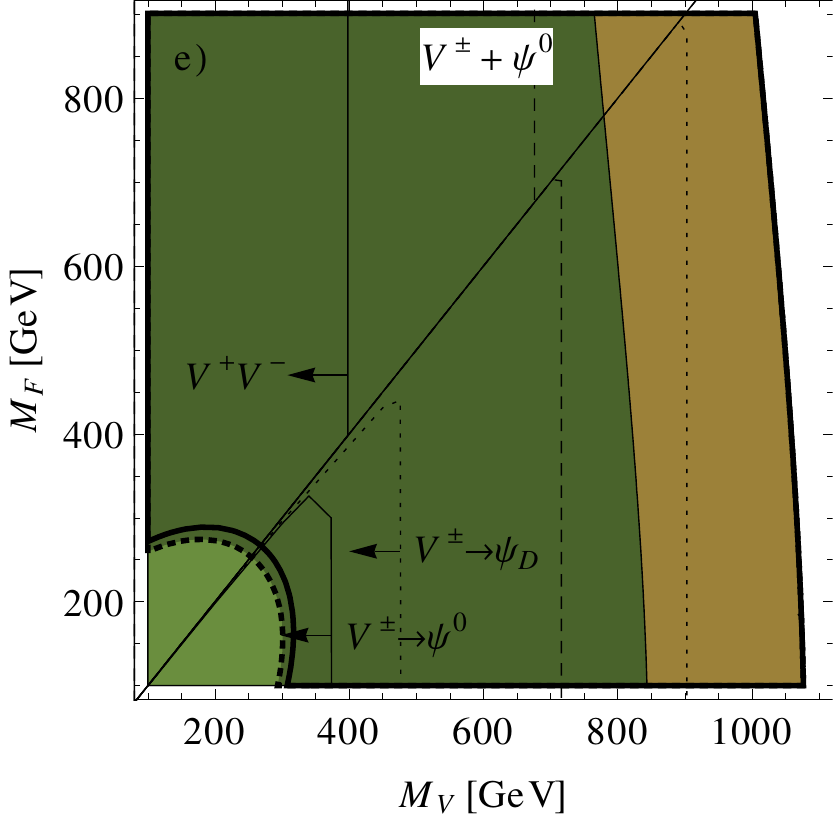}\hspace*{0.5cm} & \includegraphics[scale=0.78]{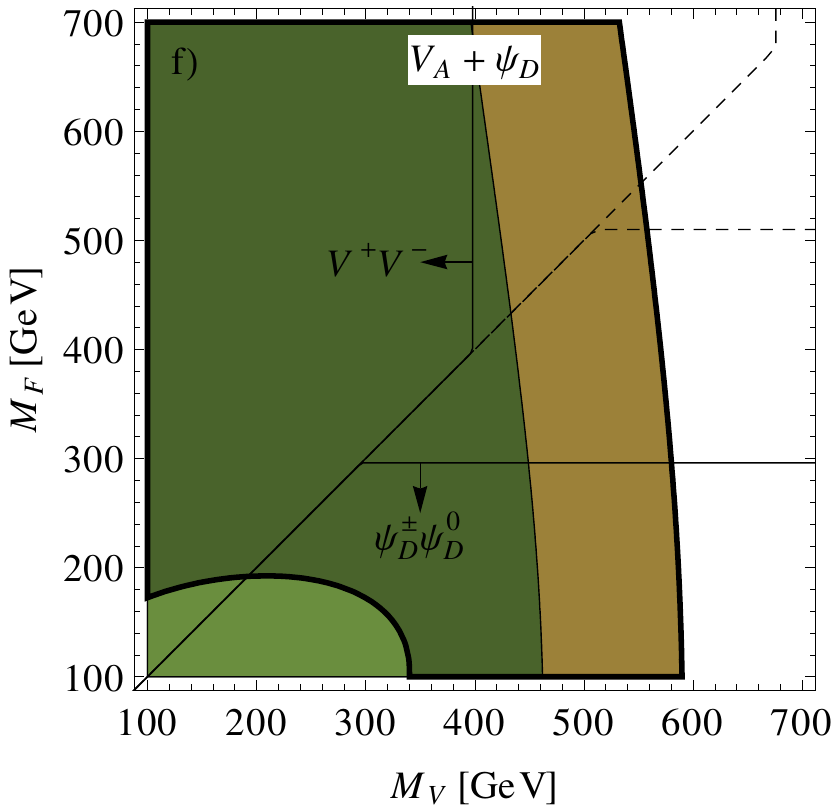}
\end{tabular}\vspace*{-0.3cm}
\caption{Contributions to $a_{\mu}$ from two new fields with different spin for $Y,g\le \sqrt{4\pi}$ (green: $1\sigma$, yellow: $2\sigma$ region). The gray area with bold boundaries is disfavored by LEP constraints on $ee\ell\ell$ contact interactions to explain $\Delta a_{\mu}$ within the $1\sigma$ range. Lower mass bounds at $95\%$ C.L. from direct searches at the 8-TeV LHC and projections for $14\tev$ (see Section~\ref{sec:lhc}) are displayed as plain and dashed black lines (dotted for $V^{\pm}+\psi_D$), respectively.}
\label{fig:diffspin}
\end{figure}

\paragraph{Neutral scalar ($\phi^0$) and charged fermion ($\psi^\pm$):} This scenario can contribute to $a_{\mu}$ through the diagram in Figure~\ref{fig:twop}~(a) with the corresponding couplings,
\begin{equation}
{\cal L} \supset -Y\overline{\ell_L}\phi^0\psi^-_R + \text{h.c.} \qquad \text{or}
\qquad {\cal L} \supset -Y\overline{\ell_R}\phi^0\psi^-_L + \text{h.c.}
\end{equation}
The former coupling applies if either of the new particles is part of an SU(2) doublet and the other one is a singlet, whereas the latter coupling is relevant if both new particles are either singlets or part of a doublet. The chirality of the SM lepton is thus determined by the electroweak properties of the new particles. The discrepancy $\Delta a_{\mu}$ in \eqref{d} can be explained at the one-(two-)sigma level for
\begin{equation}
Y > 1.9\,(1.5),\qquad M_{\phi,\psi} > 100\gev.
\end{equation}
In Figure~\ref{fig:diffspin}~(a), we display the parameter space for $\phi^0+\psi^\pm$ that accommodates $\Delta a_{\mu}$ at the one- and two- and sigma level (green and yellow areas) for $Y\le \sqrt{4\pi}$ in terms of the scalar and fermion masses $M_\phi$ and $M_\psi$. Constraints from four-lepton contact interactions are absent if $\phi^0$ is self-conjugate. In this case the box diagram in Figure~\ref{fig:box}~(a) is cancelled by a second diagram with crossed fermion lines in the final state. If the neutral scalar is part of a weak doublet, constraints from contact interactions exclude the entire parameter region for $\Delta a_{\mu}$ in Figure~\ref{fig:diffspin}~(a).

\paragraph{Charged scalar ($\phi^\pm$) and neutral fermion ($\psi^0$):} This combination of fields contributes to $a_{\mu}$ through the diagram in Figure~\ref{fig:twop}~(b) with the following couplings, 
\begin{equation}
{\cal L} \supset -Y\overline{\ell_L}\phi^-\psi^0_R + \text{h.c.} \qquad \text{or}
\qquad {\cal L} \supset -Y\overline{\ell_R}\phi^-\psi^0_L + \text{h.c.}
\end{equation}
The electroweak properties determine the structure of the coupling as in the previous case with $\phi^0+\psi^\pm$. The contribution to $a_{\mu}$ is negative and cannot explain the observed discrepancy.

\paragraph{Scalar doublet ($\phi_D$) and fermion doublet ($\psi_D$):} The two doublets defined in (\ref{eq:scalardoublet}) and (\ref{fd}) couple to right-handed SM leptons via
\begin{equation}
{\cal L} \supset -Y\overline{\psi_{D,L}}\phi_D\ell_R + \text{h.c.}
\end{equation}
The sum of contributions from the neutral and charged components of the scalar doublet, Figure~\ref{fig:twop}~(a,b), yields a positive correction to $a_{\mu}$. However, the result is too small to explain the discrepancy in \eqref{d}. Furthermore, constraints on $ee\ell\ell$ interactions at LEP exclude an explanation of $\Delta a_{\mu}$ within two sigma. Any significant contribution to $a_{\mu}$ from $\phi_D+\psi_D$ is thereby strongly disfavored.

\paragraph{Scalar doublet ($\phi_D$) and fermion adjoint triplet ($\psi_A$):} Contributions of $\phi_D+\psi_A$ to $a_{\mu}$, with $\psi_A$ defined in (\ref{fa}), correspond to the diagrams in Figure~\ref{fig:twop}~(a,b) with the coupling
\begin{equation}
{\cal L} \supset -Y\tilde{\phi}_D^\dagger\overline{\psi_{A,R}}L_L + \text{h.c.}
\end{equation}
Due to the different SU(2) structure, the (negative) contribution of $\phi_D^-+\psi_A^0$ is reduced by a factor of $(\sqrt{2})^{-4}$ with respect to the previous scenario $\phi_D+\psi_D$. A priori, the discrepancy $\Delta a_{\mu}$ can be explained at the one-(two-)sigma level for
\begin{equation}
Y > 2.7\,(2.1),\qquad M_{\phi,\psi} > 100\gev.
\end{equation}
Figure~\ref{fig:diffspin}~(b) shows the full parameter space that covers $\Delta a_{\mu}$. Four-lepton contact interactions are generated by the diagrams in Figure~\ref{fig:box}~(a,b). LEP bounds on $ee\ell\ell$ interactions (shaded gray) exclude all of the available parameter space that explains $\Delta a_{\mu}$ at the two-sigma level. In the low-mass range $M_{\phi,\psi}\sim \sqrt{s}\sim 200\gev$, constraints from $ee\ell\ell$ interactions should be taken with care, since the dynamics of the new particles beyond the zero-momentum approximation are important. For our purposes, which focus on LHC constraints, it suffices to state that effects from $\phi_D+\psi_A$ on $a_{\mu}$ are strongly suppressed, if not ruled out by LEP bounds on $ee\ell\ell$ interactions.


\paragraph{Scalar doublet ($\phi_D$) and fermion triplet ($\psi_T$):} Compared to the previous scenarios, the presence of the triplet $\psi_T$ with hypercharge $-1$, defined in (\ref{ft}), introduces new contributions to $a_{\mu}$ with doubly-charged leptons through the coupling
\begin{equation}
{\cal L} \supset -Y\phi_D^\dagger\overline{\psi_{T,R}}L_L + \text{h.c.}
\end{equation}
The corresponding diagrams are given in Figure~\ref{fig:twop}~(a,e,f). The scenario $\phi_D+\psi_T$ can explain $\Delta a_{\mu}$ at the one-(two-)sigma level for
\begin{equation}
Y > 1.0\,(0.8),\qquad M_{\phi,\psi} > 100\gev.
\end{equation}
The full parameter space is given in Figure~\ref{fig:diffspin}~(c). Constraints from $ee\ell\ell$ interactions are due to the diagrams displayed in Figure~\ref{fig:box}~(a,c). They exclude large parts (the gray area) of the parameter space for $\Delta a_{\mu}$. Potential contributions at the one-sigma level are thereby confined to a small region of the parameter space with light masses $M_{\phi,\psi}\sim 100-150\gev$.

\paragraph{Scalar adjoint triplet ($\phi_A$) and fermion doublet ($\psi_D$):} The scenario with a scalar triplet $\phi_A$ with hypercharge $0$ and a fermion doublet $\psi_D$ contributes to $a_{\mu}$ through the diagrams in Figure~\ref{fig:twop}~(a,b) with the coupling
\begin{equation}
{\cal L} \supset -Y\overline{\psi_{D,R}}\,\phi_AL_L + \text{h.c.},\qquad\qquad \phi_A = \begin{pmatrix}
\phi_A^0/\sqrt{2} & \phi_A^{+} \\ \phi_A^{-} & -\phi_A^{0}/\sqrt{2}
\end{pmatrix}.
\end{equation}
The result is negative and cannot accommodate $\Delta a_{\mu}$.

\paragraph{Scalar adjoint triplet ($\phi_A$) and fermion triplet ($\psi_T$):} Contributions to $a_{\mu}$ arise from the diagrams in Figure~\ref{fig:twop}~(a,b,e,f) through the coupling
\begin{equation}
{\cal L} \supset -Y\,\text{tr}\bigl\{ \phi_A^\dagger\overline{\psi_{T,L}}\bigr\}\ell_R.
\end{equation}
This scenario can accommodate $\Delta a_{\mu}$ in the one-(two-)sigma region with couplings
\begin{equation}
Y > 1.1\,(0.9),\qquad M_{\phi,\psi} > 100\gev.
\end{equation}
The complete parameter range with perturbative couplings is shown in Figure~\ref{fig:diffspin}~(d). Constraints from four-lepton interactions are absent due to cancellations among box diagrams with $\phi_A^0+\psi_T^-$ and among diagrams with $\phi_A^-$ and leptons $\psi_T^0,\,\psi_T^{--}$.

\paragraph{Scalar triplet ($\phi_T$) and fermion doublet ($\psi_D$):} The diagrams for $a_{\mu}$ with $\phi_T$ defined in (\ref{eq:scalartriplet}) and $\psi_D$ are given in Figure~\ref{fig:twop}, induced by the coupling
\begin{equation}
{\cal L} \supset -Y\overline{\psi_{D,L}}\phi_Ti\sigma_2L_L^c + \text{h.c.}
\end{equation}
The contribution to $a_{\mu}$ is negative and thus not appropriate to explain $\Delta a_{\mu}$.

\paragraph{Scalar triplet ($\phi_T$) and fermion adjoint triplet ($\psi_A$):} These two triplets induce corrections to $a_{\mu}$ through the diagrams in Figure~\ref{fig:twop}~(a,b,c,d) with the coupling
\begin{equation}
{\cal L} \supset -Y\,\text{tr}\bigl\{ \phi_T^\dagger\overline{\psi_{A,L}}\bigr\}\ell_R.
\end{equation}
Also in this case, the contribution to $a_{\mu}$ is negative and not able to account for the observed discrepancy.

\paragraph{Neutral vector singlet ($V^{0}$) and charged fermion ($\psi^{\pm}$):} This combination contributes to $a_{\mu}$ through the diagram in Figure~\ref{fig:twop}~(g). The fermion $\psi^{\pm}$ can be either a weak singlet or part of a doublet, which determines the coupling
\begin{equation}
{\cal L} \supset -g_R\,\overline{\psi^-_R}\gamma^{\mu}\ell^-_R V^0_{\mu} + \text{h.c.} \qquad \text{or} \qquad {\cal L} \supset -g_L\,\overline{\psi_{D,L}}\gamma^{\mu}L_L V^0_{\mu}+ \text{h.c.},
\end{equation}
respectively. The resulting contribution to $a_{\mu}$ is negative, ruling out $V^0+\psi^{\pm}$ as an explanation of $\Delta a_{\mu}$.

\paragraph{Charged vector singlet ($V^{\pm}$) and neutral fermion ($\psi^{0}$):} The Feynman diagram for $a_{\mu}$ in this scenario is given in Figure~\ref{fig:twop}~(h). Similarly to the previous case, the vector fermion $\psi^0$ can be a weak singlet or part of a doublet, yielding the chiral couplings
\begin{equation}
{\cal L} \supset -g_R\,\overline{\psi^0_R}\gamma^{\mu}\ell^-_R V^{+}_{\mu} + \text{h.c.} \qquad \text{or} \qquad {\cal L} \supset -g_L\,\overline{\psi_{D,L}}\gamma^{\mu}i\sigma_2{L}^c_L V^-_{\mu}+ \text{h.c.},
\end{equation}
respectively. The scenario $V^{\pm}+\psi^0$ can accommodate $\Delta a_{\mu}$ in the one-\mbox{(two-)}sigma region with couplings
\begin{equation}
g_{R,L} > 0.5\,(0.4),\qquad M_{V,\psi} > 100\gev,
\end{equation}
as displayed for the full parameter space in Figure~\ref{fig:diffspin}~(e). Notice that the dependence of $a_{\mu}$ on the fermion mass $M_\psi$ is very weak. Constraints from $ee\ell\ell$ contact interactions mediated by the box diagram in Figure~\ref{fig:box}~(d) exclude large parts of the parameter space (the gray area in Figure~\ref{fig:diffspin}~(e), whose plain contour corresponds to the (right-chiral) fermion singlet case; the dotted contour depicts the (left-chiral) doublet case). Since the couplings to accommodate $\Delta a_{\mu}$ with light new particles are relatively weak, $ee\ell\ell$ constraints leave open a mass range of $M_{V,\psi}\sim 100-300\gev$ to explain $\Delta a_{\mu}$ within its one-sigma limits.

\paragraph{Vector adjoint triplet ($V_A$) and fermion doublet ($\psi_D$):} This scenario combines the contributions of the previous two cases from Figure~\ref{fig:twop}~(g,h). The corresponding coupling to SM leptons is left-chiral,
\begin{equation}\label{eq:VAD}
{\cal L} \supset -g_L\,\overline{\psi_{D,L}}\gamma_{\mu}V_A^{\mu}L_L + \text{h.c.},\qquad\qquad V_A = \begin{pmatrix}
V_A^0/\sqrt{2} & V_A^{+} \\ V_A^{-} & -V_A^{0}/\sqrt{2}\end{pmatrix}.
\end{equation}
This scenario can explain $\Delta a_{\mu}$ in the one-(two-)sigma region, provided
\begin{equation}
g_L > 0.9\,(0.7),\qquad M_{V,\psi} > 100\gev.
\end{equation}
Compared to the scenario $V^{\pm}+\psi^0$, the parameter space is shifted towards lower masses, see Figure~\ref{fig:diffspin}~(e). Four-lepton contact interactions induced by the diagrams in Figure~\ref{fig:box}~(d,e) restrict the one-sigma region for $\Delta a_{\mu}$ to the mass range $M_V\sim 100-300\gev$, $M_\psi\sim 100-200\gev$.

\paragraph{Vector adjoint triplet ($V_A$) and fermion triplet ($\psi_T$):} The two triplets defined in (\ref{ft}) and (\ref{eq:VAD}) couple to muons through
\begin{equation}
{\cal L} \supset -g_R\,\overline{\psi_{T,R}}\gamma_{\mu}V_A^{\mu}\ell^-_R + \text{h.c.}
\end{equation}
and contribute to $a_{\mu}$ via the diagrams in Figure~\ref{fig:twop}~(g,h,i,j). The result is negative and cannot accommodate the discrepancy $\Delta a_{\mu}$.


\section{LHC constraints} \label{sec:lhc}

In the previous sections, the minimal new-physics scenarios that could potentially accommodate the muon magnetic moment anomaly in \eqref{d} have been identified. These are
\begin{itemize}
\item for one new field: $V^\pm$;
\item for two mixed fermion fields: $\psi_D+\psi^{\pm}$, $\psi_D+\psi^0$, $\psi_D+\psi_A$, $\psi_D+\psi_T$;
\item for two different-spin fields: $\phi^0+\psi^\pm$, $\phi_D+\psi_A$, $\phi_D+\psi_T$, $\phi_A+\psi_T$, $V^\pm+\psi^0$ and $V_A+\psi_D$.
\end{itemize}
This section is devoted to investigating how the preferred parameter space for explaining $\Delta a_\mu$ in these scenarios is constrained by current LHC data and may be further probed with the future 14-TeV run. As mentioned in Section~\ref{sec:twofields}, in some two-field cases the allowed parameter space is already severely limited by bounds on loop-induced four-lepton interactions from LEP2 (the gray regions in Figure~\ref{fig:diffspin}). However, these four-lepton corrections may conceivably be canceled by tree-level contributions from the exchange of a very heavy neutral vector boson $V^0$ (which would have a minimal effect on $a_\mu$, see Section~\ref{sec:onefield}). Therefore we will also explore the parameter space that is nominally excluded by four-lepton interactions. We do not consider the collider phenomenology for scenarios with mixing vector fermions, since the parameter range for $\Delta a_{\mu}$ in these scenarios exceeds the reach of the LHC (see Section~\ref{sec:twomixed}).

To minimize the model dependence, we focus on production of the new particles through the Drell-Yan process, which involves only gauge couplings. In particular, charged particles $X^\pm$ can be pair-produced through the partonic process $q\bar{q} \to X^+X^-$ via s-channel photon and $Z$-boson exchange. In the case of SU(2) multiplets with both charged and neutral components, one also has the associated production $q\bar{q}' \to X^\pm X^0$ via s-channel $W^\pm$ exchange.

In scenarios with two new fields, we will always look for constraints on the pair production of the lighter of the two. In this way, we circumvent cascade decays from the heavier to the lighter field, which would lead to more complex signatures. The scenarios $\phi^0+\psi^\pm$ and $V^\pm+\psi^0$ involve a new particle that is a SM gauge singlet. In this special case, Drell-Yan production of singlet pairs is not possible, so that we will instead consider cascade decays from the heavier charged particle. Due to the fact that relatively large couplings in the new-physics sector are required to explain $\Delta a_\mu$, the decay into the singlet is expected to be the dominant decay mode of the heavy charged particle.

Since the new fields need to couple to muons, we generically expect them to decay lep\-to\-ni\-cal\-ly. In addition, the possible decay modes are constrained by MFV. For a neutral scalar, $\phi^0$, these two considerations naturally imply the decay $\phi^0 \to \ell^+\ell^-$, $\ell=e,\mu,\tau$, which is universal in lepton flavor. Similarly, the characteristic decay of a charged scalar, $\phi^\pm$, is given by \mbox{$\phi^+ \to \ell^+\nu_\ell$, $\ell=e,\mu,\tau$}. The typical decays of new heavy vector bosons are completely analogous, i.e.\ $V^0 \to \ell^+\ell^-$ and $V^+\to \ell^+\nu_\ell$.
For heavy fermions, MFV mandates that they transform in the fundamental representation of the lepton flavor symmetry, so that there are three flavor copies $\psi_\ell$, $\ell=e,\mu,\tau$. The characteristic decay modes for neutral and charged fermions are given by $\psi^0 \to \nu Z,\,\,\nu H,\,\ell^- W^+$ and
 $\psi^- \to \ell^- Z,\,\ell^- H,\,\nu W^-$, respectively, with the branching fractions determined by the SU(2) representation of $\psi^{0,\pm}$ (see below). Lacking public results on LHC searches for doubly-charged fermions, we will instead constrain scenarios with triplet fermions through their neutral and singly-charged components.

Table~\ref{tab:modes} summarizes the production and decay modes considered for deriving the LHC constraints in this section.
\begin{table}
\centering
\renewcommand{\arraystretch}{1.4}
\begin{tabular}[t]{|c|l|c|c|}
\hline
Scenario & Production & LHC8 & LHC14\\
\hline
\hline
$V^\pm$ & $pp \to V^+V^-$ & $M_V > 398\gev$ & $M_V > 676\gev$ \\
\hline
$\phi^0+\psi^\pm$ 
 & $M_\psi < M_\phi$: $pp \to \psi^+\psi^-$ & -- & $\times$ \\[-.5ex]
 & $M_\psi > M_\phi$: $pp \to \psi^+\psi^- \to \ell^+ \phi^0 \,\ell^- \phi^0$ & $\times$ & $\times$ \\
\hline
$\phi^0+\psi_D$ 
 & $M_\psi < M_\phi$: $pp \to \psi^\pm\psi^0$ & $\times$ & $\times$ \\[-.5ex]
 & $M_\psi > M_\phi$: $pp \to \psi^+\psi^- \to \ell^+ \phi^0 \,\ell^- \phi^0$ & $\times$ & $\times$\\
\hline
$\phi_D+\psi^\pm$ 
 & $M_\psi < M_\phi$: $pp \to \psi^+\psi^-$ & -- & $\times$ \\[-.5ex]
 & $M_\psi > M_\phi$: $pp \to \phi^\pm\phi^0$ & $\times$ & $\times$ \\
\hline
$\phi_D+\psi_A$
 & $M_\psi < M_\phi$: $pp \to \psi^\pm\psi^0$ & $\times$ & $\times$ \\[-.5ex]
 & $M_\psi > M_\phi$: $pp \to \phi^\pm\phi^0$ & $\times$ & $\times$ \\
\hline
$\phi_{D}+\psi_T$
 & $M_\psi < M_\phi$: $pp \to \psi^\pm\psi^0$ & $M_{\psi} > 258\gev$ & $M_{\psi} > 420\gev$\\[-.5ex]
 & $M_\psi > M_\phi$: $pp \to \phi^\pm\phi^0$ & $M_{\phi} > 380\gev $ & $\times$\\
\hline
$\phi_{A}+\psi_T$
 & $M_\psi < M_\phi$: $pp \to \psi^\pm\psi^0$ & $M_{\psi} > 258\gev$ & $\times$\\[-.5ex]
 & $M_\psi > M_\phi$: $pp \to \phi^\pm\phi^0$ & $\times$ & $\times$\\
\hline
$V^\pm+\psi^0$ 
 & $M_V < M_\psi$: $pp \to V^+V^-$ & $M_V > 398\gev$ & $M_V > 676\gev$ \\[-.5ex]
 & $M_V > M_\psi$: $pp \to V^+V^- \to \ell^+ \psi^0 \, \ell^- \psi^0$ & $M_V > 373\gev$ & $M_V > 716\gev$ \\
\hline
$V^\pm+\psi_D$ 
 & $M_V < M_\psi$: $pp \to V^+V^-$ & $M_V > 398\gev$ & $M_V > 676\gev$ \\[-.5ex]
 & $M_V > M_\psi$: $pp \to V^+V^- \to \ell^+ \psi^0 \, \ell^- \psi^0$ & $M_V > 476\gev$ & $M_V > 903\gev$ \\
\hline
$V_A+\psi_D$
 & $M_V < M_\psi$: $pp \to V^+V^-$ & $M_V > 398\gev$ & $\times$ \\[-.5ex]
 & $M_V > M_\psi$: $pp \to \psi^\pm\psi^0$ & $M_{\psi} > 296\gev$ & $\times$\\
\hline
\end{tabular}\\
\vspace{1em}
\begin{tabular}[t]{|l||c@{$\,\to\,$}l|c@{$\,\to\,$}l|c@{$\,\to\,$}l|}
\hline
Decay & $\phi^0$ & $\ell^+\ell^-$ & $V^0$ & $\ell^+\ell^-$ & $\psi^0$ & $\nu Z,\,\,\nu H,\,\ell^\pm W^\mp$ \\[-.2ex]
 & $\phi^\pm$ & $\ell^\pm \nu$ & $V^\pm$ & $\ell^\pm \nu$ & $\psi^\pm$ & $\ell^\pm Z,\,\ell^\pm H,\,\nu W^\pm$ \\
\hline
\end{tabular}
\caption{LHC production (top) and typical decay process (bottom) for the new particles in the one- and two-field scenarios that can explain the muon magnetic moment anomaly. Cases that are excluded at two sigma by 8-TeV LHC data or can be probed conclusively at 14 TeV are marked by a cross. Wherever the two-sigma range of $\Delta a_{\mu}$ is not fully covered, we display the lower mass bounds as obtained from the analyses described in the text.}
\label{tab:modes}
\end{table}
 For concreteness, we will assume that there are no additional decay modes besides those listed in the table. For the new heavy scalar and vector bosons, MFV would in principle also permit decay channels into quarks, SM weak gauge bosons, or Higgs bosons. Furthermore, there may be exotic decays into additional light states of the new-physics sector that do not play any role for $a_\mu$. Therefore the reader should bear in mind that the presence of any decay channels beyond those listed in Table~\ref{tab:modes} would reduce the observable signal at the LHC and thus weaken the limits presented below.

\subsection{Constraints from existing 8-TeV LHC data}

To derive the constraints on the viable parameter space of our simplified scenarios from existing LHC data, we use results published by the ATLAS and CMS collaborations for new-physics searches in particular models, and recast them to the processes considered here. The resulting bounds on the masses of new particles are illustrated in Figures~\ref{fig:chargedvector} and \ref{fig:diffspin}.

\paragraph{$\bullet$ $pp \to \phi^\pm \phi^0 \to \ell^\pm\nu_\ell
\ell'^+\ell'^-$:}{ This process can be constrained using results of a
search for supersymmetric charginos and neutralinos by ATLAS based on a
signature with three leptons and missing energy \cite{13035} (for a
similar analysis by CMS, see \cite{sus13006}). The strongest limits are
obtained in the signal region referred to as SRnoZc in \cite{13035}. We
have used {\sc CalcHEP} to compute the signal rate in our scenario,
implementing these cuts together with basic selection cuts from
\cite{13035}. We assume that the scalars decay into the three
generations of SM leptons with equal probability and there are no other
decay channels. The mass bound was determined by finding the
mass which generated the 95\% C.L. upper limit on the signal cross section as given in Table
4 of \cite{13035}.

We find that the current ATLAS data sets a bound on the mass of a scalar doublet,
\mbox{$M_{\phi_D} > 395\gev$} at 95\% C.L. This eliminates all allowed parameter
space of $\Delta a_{\mu}$ for $\phi_D+\psi^{\pm}$ and $\phi_D+\psi_A$
(both for $M_\phi < M_\psi$),
and part of the allowed parameter space for $\phi_D+\psi_T$ ($M_\phi <
M_\psi$). The bound for a scalar weak triplet is $M_{\phi_A} > 456\gev$ at
95\% C.L. Due to the isospin-enhanced coupling to gauge bosons, the
constraint is stronger than for the doublet. It excludes the entire
parameter space of $\Delta a_{\mu}$ in the scenario $\phi_A+\psi_T$ for
$M_{\phi} < M_{\psi}$.}

\paragraph{$\bullet$ $pp \to \psi^\pm \psi^0 \to Z\ell^\pm W^\pm\ell^\mp \to \ell'^+\ell'^-\ell^\pm W^\pm\ell^\mp$:}{This process is very similar to pair production of heavy fermions in the type-III seesaw model. Limits on this model have been obtained by ATLAS \cite{13009} and CMS \cite{exo11073}. Here the ATLAS analysis has been used to put limits on the production of weak doublet and triplet vector fermions. The cross sections for $pp \to \psi^\pm \psi^0$ were computed in {\sc CalcHEP}, assuming that the vector fermions are lepton flavor triplets, as mandated by MFV. Since the experimental searches are sensitive to both electrons and muons, this leads to a factor of two for the production rate. The computed numbers for cross section times branching ratio were compared to the observed 95\% C.L.\ line in Figure~3 of \cite{13009}.

For triplet fermions, the branching ratios are given by $\mathcal{B}(\psi^\pm \to Z\ell^\pm) = 1/4$ and \mbox{$\mathcal{B}(\psi^0
\to  W^{\pm}\ell^{\mp}) = 1/2$}, which leads to the limit $M_{\psi_{A,T}} > 258\gev$. Doublet fermions have a smaller production cross section, but larger branching ratios \mbox{$\mathcal{B}(\psi^\pm \to Z\ell^\pm) = 1/2$} and $\mathcal{B}(\psi^0 \to W^{\pm}\ell^{\mp}) = 1$,
resulting in the limit $M_{\psi_D} > 296\gev$. For the cases with a new fermion and a new scalar field, these bounds
eliminate all allowed parameter space for $\phi_D+\psi_A$ and part of the parameter space for $\phi_D+\psi_T$ and $\phi_A+\psi_T$ (all for $M_\psi < M_\phi$). Similarly, they exclude part of the viable parameter region for $V_A+\psi_D$
(for $M_\psi < M_V$).

\paragraph{$\bullet$ $pp \to \psi^\pm \psi^\mp \to Z\ell^\pm Z/H\ell^\mp \to \ell'^+\ell'^-\ell^\pm \ell^\mp + {\rm hadrons}$:}{For charged singlet fermions, the process described in the previous item does not exist. However, if one fermion in $\psi^+\psi^-$ decays into a $Z$ boson, while the other one decays into a $Z$ or Higgs boson, one obtains a very similar final-state signature with four leptons, two of which reconstruct the $Z$ invariant mass. Therefore, the cross-section bounds from \cite{13009} can be applied approximately also to this case. We assume that the second $Z$ boson decays non-leptonically to account for the second $Z$ veto in the ATLAS analysis. Computing signal cross sections with {\sc CalcHEP} as above and folding in the branching fractions $\mathcal{B}(\psi^\pm \to Z\ell^\pm) = \mathcal{B}(\psi^\pm \to H\ell^\pm) = 1/4$, we find that no limit can be placed on singlet fermion pair production with the result of \cite{13009}. This mainly follows from the fact that the production cross section for $\psi^+\psi^-$, which have only hypercharge but no weak isospin, is suppressed due to the relatively small hypercharges of the initial-state quarks.}

\paragraph{$\bullet$ $pp \to V^+V^- \to \ell^+\ell'^-\nu_\ell\overline{\nu}_{\ell'}$:}{This process can be constrained from searches for slepton pair production, where
each slepton decays into a charged lepton and a neutralino \mbox{\cite{sus13006,13049}}. To translate the slepton limits to vector boson
pair production, the
cross sections for \mbox{$pp \to V^+V^-$} were computed with {\sc CalcHEP}, assuming a
branching fraction of 1/3 each into $\ell=e$ and $\ell=\mu$ (the remaining third
for $\ell=\tau$ is not used in the experimental analyses). The results 
were compared to the 95\% C.L.\ upper bounds in Figure~20~(right) in \cite{sus13006} in the case where the neutralino mass is set to
zero. With this procedure, the lower limit on the vector boson mass, $M_V > 398\gev$, is obtained. This mass bound rules out a portion of the allowed parameter space for $V^\pm$, $V^\pm+\psi^0$ and $V_A+\psi_D$ (for $M_V < M_\psi$).}

\paragraph{$\bullet$\hspace*{-0.2cm}}{$pp \to V^+V^- \to \ell^+ \psi^0\ell^- \psi^0$ for $V^\pm+\psi^0$: With further decays $\psi^0 \to W^\pm\, \ell^\mp$, this process leads to a four-lepton signal. Thus, the masses of $V$ and $\psi^0$ can be
constrained from an ATLAS search \cite{13009}, which considers events with
four or more charged leptons ($e,\mu$) in the final state. Using {\sc CalcHEP} we computed the signal rate including basic selection cuts
as described in \cite{13009}. This signal rate was added to the SM
background and limits were determined through comparison with the observed
event yield (background and observations are given in the top row of Table~2 in \cite{13009}).

If $\psi^0$ is part of a weak doublet, the branching ratio is $\mathcal{B}(\psi^0\rightarrow W^\pm\ell^\mp) = 1$. We obtain the limit $M_V > 476\gev$,
provided $M_{\psi}$ is sufficiently smaller than $M_V$. For $M_{\psi}\lesim M_V$, the decay produces
soft leptons, which do not pass the detector cuts. As a result, there is a small gap in the excluded 
parameter space (see Figure~\ref{fig:diffspin}~(e)) near the line of $M_V=M_{\psi}$. The width of the mass gap is $19 \gev$ for $M_V = 451\gev$ and shrinks to less than
$4\gev$ for $M_V < 300\gev$. If $\psi^0$ is a weak singlet, the branching ratio is reduced to $\mathcal{B}(\psi^0\rightarrow W^\pm\ell^\mp) = 1/2$. We obtain the less stringent limit $M_V > 373\gev$, again
assuming that $M_{\psi}$ is sufficiently smaller than $M_V$. The mass gap
is $14 \gev$ for $M_V = 340\gev$ and shrinks to less than $2\gev$ for $M_V
< 200\gev$. This excludes part of the allowed parameter space for the scenarios $V^\pm+\psi_D$ and $V^\pm+\psi^0$ (for
$M_V > M_\psi$).

\paragraph{$\bullet$\hspace*{-0.2cm}}{$pp \to \psi^+\psi^- \to \ell^+ \phi^0\ell^- \phi^0$ for $\phi^0+\psi^{\pm}$:\ \ This cascade with the subsequent decay $\phi^0\rightarrow \ell^+ \ell^-$ is relevant if both the fermion and the (lighter) scalar are weak singlets. We recast the ana\-ly\-sis of $pp \to V^+V^- \to \ell^+ \psi^0\ell^- \psi^0$ described above for $\phi^0+\psi^{\pm}$ by adapting the production cross section to a pair of charged fermions. The kinematics of the first decay steps are similar in both scenarios, while the decay of the scalar $\phi^0$ typically yields more leptons in the final state compared to the fermion $\psi^0$. Therefore we obtain conservative limits if we assume that the event yield passing the detector cuts is similar in both scenarios. The resulting bound on the fermion mass is $M_{\psi} > 405\gev$ at $95\%$ C.L. This excludes the entire parameter region for $\Delta a_{\mu}$ in the two-singlet scenario $\phi^0+\psi^{\pm}$ with $M_{\psi} > M_{\phi} + 5\gev$. Since the cross section for a pair of charged doublet fermions is about a factor of two larger than for singlet fermions, the same analysis also excludes the scenario $\phi^0+\psi_D$ ($M_{\psi_D} > M_{\phi}$) as a possible explanation of $\Delta a_{\mu}$.}\\

The mass bounds obtained for each scenario with 8-TeV data are listed in Table~\ref{tab:modes}. Excluded (unconstrained) scenarios are marked by a cross (a hyphen). As is apparent from the table, the scenarios $\phi_D+\psi_D$ and $\phi_D+\psi_A$ are already excluded at the two-sigma level by LHC searches. Taking LEP constraints from one-loop $ee\ell\ell$ contact terms into account, all scenarios are excluded but those with a neutral or weak adjoint scalar, where contributions to $ee\ell\ell$ interactions cancel. In some scenarios, especially those with new vector bosons, the viable parameter space reaches out to mass scales in the TeV range. As we will show in the following section, the higher collision energy at the 14-TeV LHC will be beneficial to test those high-mass regions.

\subsection{Projections for the 14-TeV LHC}

For the 14-TeV projections, we follow the strategy of \cite{cmsproj}. Starting from the existing \mbox{8-TeV} searches by ATLAS and CMS (referenced in the previous subsection), the expected event yields were obtained by scaling the luminosity to 300~fb$^{-1}$ and multiplying with the ratio of cross sections $\sigma_{\rm sig(bkg)}(14\tev)/\sigma_{\rm sig(bkg)}(8\tev)$. The total production cross section $\sigma_{\rm sig(bkg)}(\sqrt{s})$ for the signal (dominant backgrounds) at the $pp$ CM energy of $\sqrt{s}$ was computed with {\sc CalcHEP}. This approach assumes that the selection efficiency for the signal and background will remain similar when going from an 8-TeV to a 14-TeV analysis. While this assumption is admittedly rather ad hoc, a more refined estimation would require a full-fledged simulation, which is beyond the scope of this paper. 
Since the signal cross section varies very rapidly as a function of the produced particles' masses, we believe that our projected mass limits will not be strongly influenced by the uncertainties in the selection efficiency and thus should give a meaningful indication of the reach of the 14-TeV LHC. Furthermore, several of the existing ATLAS and CMS analyses used above are not optimized for our new-physics signatures, so that we expect our projected bounds to be rather conservative.

Using this procedure to re-scale the analyses of the previous subsection, we obtain the following expected exclusion limits for the 14-TeV LHC:

\paragraph{$\bullet$ $pp \to \phi^\pm \phi^0 \to \ell^\pm\nu_\ell \ell'^+\ell'^-$:}{For scalar doublets, we obtain the projected mass bound of \mbox{$M_{\phi_D} > 660\gev$}. If no signal is observed, this will rule out the entire parameter space for $\Delta a_{\mu}$ in the scenario $\phi_D+\psi_T$ for $M_\phi < M_{\psi}$. The projection for the scalar adjoint triplet pushes the mass limit up to $M_{\phi_A}>760\gev$.}

\paragraph{$\bullet$ $pp \to \psi^\pm \psi^0 \to Z\ell^\pm W^\pm\ell^\mp \to \ell'^+\ell'^-\ell^\pm W^\pm\ell^\mp$:}{For triplet fermions, the projected mass bound is $M_{\psi_{A,T}} > 420\gev$, while for doublet fermions we obtain $M_{\psi_D} > 510\gev$. These estimates probe the entire parameter region for $\phi_A+\psi_T$ and almost the complete region for $\phi_D+\psi_T$ and $V_A+\psi_D$ (all for $M_{\psi}<M_{\phi,V}$).}

\paragraph{$\bullet$ $pp \to \psi^\pm \psi^\mp \to Z\ell^\pm Z/H\ell^\mp \to \ell'^+\ell'^-\ell^\pm \ell^\mp + {\rm hadrons}$:}{The increased luminosity and production energy at the 14-TeV LHC allow us to set a first lower bound on the mass of electroweak singlet fermions, $M_{\psi} > 240\gev$. It covers the full parameter space of $\Delta a_{\mu}$ for $\phi^0+\psi^{\pm}$ and $\phi_D+\psi^{\pm}$ (both for $M_{\psi} < M_{\phi}$).}

\paragraph{$\bullet$ $pp \to V^+V^- \to \ell^+\ell'^-\nu_\ell\overline{\nu}_{\ell'}$:}{The projected bound for the production of two new vector fermions is $M_V > 676\gev$. This will probe the full parameter space of $\Delta a_{\mu}$ in the scenario $V_A+\psi_D$ for $M_V < M_{\psi}$ and a significant portion of parameter space in the scenarios $V^{\pm}$ and $V^\pm+\psi^0$ (for $M_V < M_{\psi}$). }

\paragraph{$\bullet$ $pp \to V^+V^- \to \ell^+ \psi^0\ell^- \psi^0$:}{The projected mass limits reach $M_V > 716\gev$ for a singlet fermion and $M_V > 903\gev$ for a doublet fermion (both for $M_V > M_{\psi}$). This corresponds to part of the parameter space for the scenarios $V^{\pm} + \psi^0$ and $V^{\pm} + \psi_D$.}\\

The limits on the parameter space of each specific scenario are marked in Figures~\ref{fig:chargedvector} and~\ref{fig:diffspin} as dashed lines. From the plots and from our summary in Table~\ref{tab:modes}, it is apparent that the 14-TeV LHC has a strong potential to conclusively probe most viable scenarios for $\Delta a_{\mu}$. All scenarios with new scalars and a vector boson triplet can be tested (the small open corner of parameter space for $\phi_D+\phi_T$ will presumably be closed with refined analyses). In scenarios with a singlet vector boson, the 14-TeV data can push the mass bounds to regions of parameter space where strong couplings $g_R \gesim 3.0$ or $g_L \gesim 3.8$ to leptons are required to explain $\Delta a_{\mu}$ at two sigma. These regions, however, are already excluded by LEP searches for four-lepton contact interactions, unless those constraints are relaxed by additional fields in a specific model. Combining LEP and 14-TeV LHC data, all of the minimal models considered in this work can thus be either excluded or conclusively tested.


\section{\boldmath $\tan\beta$-enhanced corrections} \label{sec:tanb}

In Sections~\ref{sec:onefield}--\ref{sec:twofields} we found that a weakly coupled new-physics explanation for the $a_\mu$ discre\-pancy requires that at least some of the new particles have masses of a few 100~GeV, with upper 95\%~C.L.\ bounds typically significantly below 1~TeV. As a result, the LHC can search for these particles in a fairly model-independent way, as we discussed in the previous section.

However, in some models the correction to $a_\mu$ can be enhanced by a factor $\tan\beta \gg 1$, where \mbox{$\tan\beta = v_2/v_1$} is the ratio of the vevs of two Higgs doublets. The best-known example of this kind is the Minimal Supersymmetric Standard Model (MSSM) \cite{mssm,mssm2}. In order to realize $\tan\beta$-enhanced contributions to $a_\mu$, the new-physics sector has to fulfill a number of conditions: 
\begin{itemize}
\item It needs to contain a second Higgs doublet. The muon receives its mass
from coupling to the Higgs doublet with the smaller vev, $m_\mu = y_\mu v_1/\sqrt{2}$. The Yukawa coupling \mbox{$y_\mu = \sqrt{2}m_\mu/(v \cos\beta) \approx \sqrt{2}m_\mu \tan\beta/v$}
is thus enhanced by $\tan\beta$, which leads to the corresponding enhancement of the $a_\mu$ correction.
\item There must be additional terms that break the chiral symmetry of the leptons. In the MSSM this role is played by the $\mu$ term in the superpotential.
\item The relevant one-loop diagrams should contain one $\tan\beta$-enhanced coupling proportional to $y_\mu$ (in accordance with MFV). The other couplings in the diagram should be of weak strength (i.e.\ not involving additional small muon Yukawa couplings). Typically this requires mixing between several new particles, such as gaugino--higgsino mixing or L-sfermion--R-sfermion mixing in the MSSM.
\end{itemize}
For the example of the MSSM, analytic expressions for $\delta a_\mu$ can be found for instance in \cite{mssm2}. Taking values of $\tan\beta$ in the range $30 \lesim \tan\beta \lesim 100$, the observed discrepancy $\Delta a_\mu$ in \eqref{d} can be accommodated in the MSSM even if the masses of the particles in the loop are of ${\cal O}(1\tev)$. Owing to these large masses, it becomes more difficult to conclusively test this scenario at the LHC.

On the other hand, the MSSM (or any other model that can produce $\tan\beta$-enhanced corrections to $a_\mu$) is clearly more complex than the scenarios discussed in the previous sections of this paper, since it requires the introduction of four or more fields beyond the SM (the second Higgs doublet, and a boson and two mixing fermion fields in the loop, or a fermion and two mixing boson fields in the loop). This added complexity leads to a richer phenomenology and potential new signatures at the LHC, which require a dedicated (and more model-dependent) analysis. We refer the reader to the pertinent literature for the MSSM \cite{mssmlhc}, where these questions have been studied in detail.


\section{Conclusions} \label{sec:concl}

The goal of this work was to determine to what extent an explanation of the $a_{\mu}$ anomaly in terms of new particles around the electroweak scale can be probed with existing and expected data at the LHC. We have followed a model-independent approach and investigated perturbative scenarios with one or two new fields with spin and weak isospin up to one. Throughout this work, we have assumed that lepton flavor violation in the couplings of those new fields is minimal, in the sense of introducing no new sources of flavor violation besides the lepton Yukawa couplings in the SM. The assumption of MFV protects the process $\mu\rightarrow e\gamma$ from overly large effects, as discussed in Section~\ref{sec:amu-meg-mmu}. It requires that new vector leptons transform as the fundamental representation of the flavor group, which has consequences on their production and decay rates at the LHC. MFV also affects constraints from $e^+e^-$ collisions at LEP, which are based on flavor-universal couplings of new vector and scalar bosons to leptons.

In a first step, we have identified those models which can explain the discrepancy $\Delta a_{\mu}$ within its two-sigma range. A number of cases yield negative contributions to $a_{\mu}$ or are too small to explain $\Delta a_{\mu}$ with perturbative couplings. This is true in particular for all scenarios with one new vector lepton weak singlet or triplet and for a scalar triplet, prominent from neutrino mass models of seesaw-type II.

A-priori viable models with one new field are generally strongly constrained by LEP measurements (discussed in Section~\ref{sec:onefield}). Searches for resonances in $e^+e^-\rightarrow \ell^+\ell^-$ interactions exclude neutral vector bosons $V^0$, often dubbed $Z'$ bosons in a variety of models, and scalar doublets $\phi_D$, present in models with extended Higgs sectors, as possible explanations of $\Delta a_{\mu}$. Precision measurements of observables at the $Z$ pole set tight limits on the coupling of SM leptons to new vector leptons. This strongly constrains contributions to $a_\mu$ in all models with fermion fields. The only viable one-field solution to $\Delta a_{\mu}$ after LEP is a charged vector boson $V^{\pm}$ with right-chiral couplings to leptons.

Two vector leptons mixing through a Yukawa coupling $Y_{12}$ can explain $\Delta a_{\mu}$, if the mixing is sizeable (see Section~\ref{sec:twomixed}). Within MFV, a scenario where both vector leptons are in the same representation of the flavor group leaves the mixing $Y_{12}$ unconstrained, while the coupling of one of the fields to SM leptons is suppressed by the lepton Yukawa coupling, $Y_1\sim y_{\ell}$ or $Y_2\sim y_{\ell}$. LEP bounds on $Y_1$ and $Y_2$ are stronger than MFV suppression in the parameter region $M\lesim 2\tev$ for $a_{\mu}$, while MFV suppression is dominant for $a_e$. Due to the scaling relation $a_e/a_{\mu}=m_e^2/m_{\mu}^2$, MFV bounds on vector leptons in $a_e$ thus exclude the low-mass range for $\Delta a_{\mu}$. Since LEP constraints weaken as the heavy vector leptons decouple from the SM, effects of mixing vector leptons on $a_{\mu}$ may be large even for masses in the multi-TeV scale. Such a scenario can therefore not be ruled out at the 14-TeV LHC.

Models with two new fields with different spins are generally less constrained by indirect observables than the previous cases. Still, the coupling of two new fields to leptons can be significantly limited by LEP data through one-loop effects on four-lepton contact interactions. In Section~\ref{sec:twofields}, we found that these constraints exclude large parts of the viable parameter space for $a_{\mu}$ in most scenarios. As far as we know, model-independent constraints from loop-level effects on four-lepton interactions have not been established before. Our results, summarized in Appendix~\ref{sec:4lci}, may serve as a new general tool to set bounds on the coupling of one lepton to two new weakly-coupling fields in a specific model. Since one-loop effects in four-lepton interactions may be compensated for by another heavy field contributing at tree level, we consider these LEP bounds optional and less rigorous than the bounds from direct searches.

In order to test the remaining viable scenarios at the LHC, we have re-interpreted existing 8-TeV searches for fields that lead to similar signatures (see Section~\ref{sec:lhc}). They are mostly based on pair production of the relevant new particles, which subsequently decay into a final state with multiple leptons. We have evaluated the expected event yield with parton-level simulations, assuming that the decay proceeds mainly through the couplings relevant for $a_{\mu}$ and that no further exotic decay channels play a role. In some scenarios with two new fields, we additionally study cascade decays of the heavier new particle into the lighter one, which probe regions of the parameter space that are inaccessible through direct production. All possible models not excluded by indirect observables are summarized in Table~\ref{tab:modes}, together with the production and decay modes we have used to constrain the parameter space for $a_{\mu}$.

The resulting mass bounds are also listed in Table~\ref{tab:modes} and illustrated in Figures~\ref{fig:chargedvector} and~\ref{fig:diffspin}. Some scenarios are already entirely excluded by 8-TeV data, while for others the viable parameter range is pushed to high masses. Taking loop-induced LEP bounds at face value, the only remaining scenarios are those with a neutral or weak adjoint scalar, where effects on four-lepton interactions cancel. Confining ourselves to robust direct bounds, a number of models, especially those with new vector bosons, cannot be ruled out with 8-TeV data and require further investigation at the 14-TeV LHC. We have thus extrapolated our results with 8-TeV data to the 14-TeV run by rescaling the production cross section and assuming similar event yields. From Table~\ref{tab:modes}, it is apparent that the LHC has the potential to conclusively probe all scenarios with new scalars as a possible explanation of $\Delta a_{\mu}$ in its 14-TeV run. Models with new vector bosons will, if no discovery is made, be confined to strong couplings and masses around the TeV scale. In order to cover the remaining parameter space within these models, the current analyses may be refined with tailored cuts and the reconstruction of intermediate particles (for a recent approach to reconstruction in the presence of invisible decay products, see for instance \cite{Han:2009ss}).

Beyond our framework of simple models and MFV, solutions to $a_{\mu}$ exist in models with a more complicated structure, such as the MSSM discussed in Section~\ref{sec:tanb}. With our model-independent analysis, we provide a guideline for future tests of possible explanations of the $a_{\mu}$ anomaly at the LHC, and a convenient reference to estimate constraints from $a_{\mu}$ on specific similar models.


\section{Acknowledgments}
We thank Michele Redi for pointing out the MFV suppression of effects from mixing vector leptons in $a_e$. This work was supported in part by the National Science Foundation, grant PHY-1212635. Fermilab is operated by Fermi Research Alliance, LLC, under Contract No.\ DE-AC02-07CH11359 with the United States Department of Energy.


\appendix

\section{New-physics contributions to $a_{\mu}$}
\label{sec:olf}

In this appendix, we list the one-loop results for $a_{\mu}$ from contributions of one or two of the new fields defined in Table~\ref{tab:notation}. They can be expressed in terms of the loop functions
\begin{align}\label{eq:loopfunctions}
\begin{aligned}
F_{\rm FFV}(x) &= \tfrac{1}{6(x-1)^4} \bigl [-5x^4+14x^3-39x^2+38x-8+18x^2 \ln x\bigr ], \\
G_{\rm FFV}(x) &= \tfrac{1}{(x-1)^3} \bigl [x^3+3x-4-6x\ln x\bigr ], \\
F_{\rm VVF}(x) &= \tfrac{1}{6(x-1)^4} \bigl [4x^4-49x^3+78x^2-43x+10+18x^3 \ln x\bigr ],\\
G_{\rm VVF}(x) &= \tfrac{1}{(x-1)^3} \bigl [-x^3+12x^2-15x+4-6x^2 \ln x\bigr ], \\
F_{\rm FFS}(x) &= \tfrac{1}{6(x-1)^4} \bigl [x^3-6x^2+3x+2+6x \ln x\bigr ], \\
G_{\rm FFS}(x) &= \tfrac{1}{(x-1)^3} \bigl [x^2 - 4x + 3 + 2 \ln x\bigr ], \\
H_{\rm FFS}(x) &= x[F_{\rm FFS}(x)+G_{\rm FFS}(x)], \\
F_{\rm SSF}(x) &= \tfrac{1}{6(x-1)^4} \bigl [-2x^3-3x^2+6x-1+6x^2 \ln x\bigr ].
\end{aligned}
\end{align}
The results for one new field and two new fields with different spin in the loop are given in Tables~\ref{tab:onep} and \ref{tab:twop}, respectively. For new vector fermions, we retain only the leading contributions of $\mathcal{O}(\epsilon^2)$, where $\epsilon=Yv/M$ parametrizes the mixing between SM leptons and vector leptons.

\begin{table}[!h]
\centering
\renewcommand{\arraystretch}{1.6}
\begin{tabular}{|l|c|}
\hline
Neutral vector boson ($V^0$) & $\frac{m_\mu^2(3g_Lg_R-g_L^2-g_R^2)}{12\pi^2M_V^2}$ \\
\hline
Charged vector boson ($V^\pm$) & $\frac{5m_\mu^2g_R^2}{48\pi^2 M_V^2}$ \\
\hline
Scalar doublet ($\phi_D$) & $\frac{m_\mu^2 Y^2}{32\pi^2 M_\phi^2}$ \\
\hline
Scalar triplet ($\phi_T$) & $-\frac{3m_\mu^2 Y^2}{64\pi^2 M_\phi^2}$ \\
\hline
Neutral vector fermion ($\psi^0$) & 
$\frac{G_Fm_\mu^2\epsilon^2}{24\sqrt{2}\pi^2} \bigl[ -5 +
  3F_{\rm VVF}(M_\psi^2/M_W^2) \bigr]$ \\
\hline
Charged vector fermion ($\psi^\pm$) & 
 \parbox{4.1in}{$\frac{G_Fm_\mu^2\epsilon^2}{16\sqrt{2}\pi^2} \bigl[-\frac{8}{3}c_W^2 + 2 + F_{\rm FFV}(M_{\psi}^2/M_Z^2)
 + H_{\rm FFS}(M_{\psi}^2/M_H^2)\bigr\} \bigr]$}\\
\hline
Vector fermion doublet ($\psi_D$) & 
 \parbox{4.1in}{$\frac{G_Fm_\mu^2\epsilon^2}{16\sqrt{2}\pi^2}
  \bigl[\frac{8}{3}c_W^2 + \frac{4}{3} + F_{\rm FFV}(M_\psi^2/M_Z^2) + H_{\rm FFS}(M_{\psi}^2/M_H^2)
 \newline \rule{0mm}{0mm}\hfill + 2 F_{\rm VVF}(M_\psi^2/M_W^2) + 2 G_{\rm VVF}(M_\psi^2/M_W^2) \bigr]$} \\
\hline
Vector fermion triplet ($\psi_A$) & 
 \parbox{4.1in}{$\frac{G_Fm_\mu^2\epsilon^2}{16\sqrt{2}\pi^2}
  \bigl[\frac{8}{3}c_W^2 - \frac{11}{3} + F_{\rm FFV}(M_\psi^2/M_Z^2) + 2G_{\rm FFV}(M_\psi^2/M_Z^2)
 \newline \rule{0mm}{0mm}\hfill + H_{\rm FFS}(M_{\psi}^2/M_H^2) + F_{\rm VVF}(M_\psi^2/M_W^2) + 2G_{\rm VVF}(M_\psi^2/M_W^2) \bigr]$} \\
\hline
Vector fermion triplet ($\psi_T$) & 
 \parbox{4.1in}{$\frac{G_Fm_\mu^2\epsilon^2}{32\sqrt{2}\pi^2}
 \bigl[-\frac{8}{3}c_W^2 - 18 + F_{\rm FFV}(M_\psi^2/M_Z^2)
  + H_{\rm FFS}(M_{\psi}^2/M_H^2)
 \newline \rule{0mm}{0mm}\hfill + 12F_{\rm VVF}(M_\psi^2/M_W^2) + 4G_{\rm VVF}(M_\psi^2/M_W^2)
 \newline \rule{0mm}{0mm}\hfill + 8F_{\rm FFV}(M_\psi^2/M_W^2) + 8G_{\rm FFV}(M_\psi^2/M_W^2) \bigr]$} \\
\hline
\end{tabular}
\caption{Correction $\delta a_\mu$ to the muon anomalous magnetic moment from
one new field in the vertex loop. The functions $F_{\rm XYZ}$, $G_{\rm XYZ}$ and $H_{\rm XYZ}$ are defined in (\ref{eq:loopfunctions}). The notation follows the one introduced in Section~\ref{sec:onefield}.}
\label{tab:onep}
\end{table}
\begin{table}[!ht]
\centering
\renewcommand{\arraystretch}{1.5}
\begin{tabular}{|l|l|c|l|}
\hline
$\phi^0 + \psi^\pm$ & Fig.~\ref{fig:twop}~(a) &
{$\frac{m_\mu^2Y^2}{16\pi^2M_\phi^2}F_{\rm FFS}(M_\psi^2/M_\phi^2)$} & $\delta a_{\mu} > 0$ \\
\hline
$\phi^\pm + \psi^0$ & Fig.~\ref{fig:twop}~(b) &
{$\frac{m_\mu^2Y^2}{16\pi^2M_\phi^2}F_{\rm SSF}(M_\psi^2/M_\phi^2)$} & $\delta a_{\mu} < 0$\\
\hline
$\phi_{D} + \psi_{D}$ &  Fig.~\ref{fig:twop}~(a,b) & 
$\frac{m_\mu^2Y^2}{16\pi^2M_\phi^2}\bigl[F_{\rm FFS}(M_\psi^2/M_\phi^2)
  + F_{\rm SSF}(M_\psi^2/M_\phi^2) \bigr]$ & $\delta a_{\mu} > 0$ \\
\hline
$\phi_{D} + \psi_{A}$ & Fig.~\ref{fig:twop}~(a,b) & 
$\frac{m_\mu^2Y^2}{32\pi^2M_\phi^2}\bigl[2F_{\rm FFS}(M_\psi^2/M_\phi^2)
  + F_{\rm SSF}(M_\psi^2/M_\phi^2) \bigr]$ & $\delta a_{\mu} > 0$ \\ 
\hline
$\phi_{D} + \psi_{T}$ & Fig.~\ref{fig:twop}~(a,e,f) & 
$\frac{m_\mu^2Y^2}{32\pi^2M_\phi^2}\bigl[5F_{\rm FFS}(M_\psi^2/M_\phi^2)
  - 2F_{\rm SSF}(M_\psi^2/M_\phi^2) \bigr]$ & $\delta a_{\mu} > 0$ \\
\hline
$\phi_{A} + \psi_{D}$ & Fig.~\ref{fig:twop}~(a,b) & 
$\frac{m_\mu^2Y^2}{32\pi^2M_\phi^2}\bigl[F_{\rm FFS}(M_\psi^2/M_\phi^2)
  + 2F_{\rm SSF}(M_\psi^2/M_\phi^2) \bigr]$ & $\delta a_{\mu} < 0$ \\
\hline 
{$\phi_{A} + \psi_{T}$} & Fig.~\ref{fig:twop}~(a,b,e,f) & 
$\frac{m_\mu^2Y^2}{16\pi^2M_\phi^2}3F_{\rm FFS}(M_\psi^2/M_\phi^2)$ & $\delta a_{\mu} > 0$ \\
\hline
$\phi_{T} + \psi_{D}$ & Fig.~\ref{fig:twop}~(b,c,d) & 
$\frac{m_\mu^2Y^2}{32\pi^2M_\phi^2}\bigl[-2F_{\rm FFS}(M_\psi^2/M_\phi^2)
  + 5F_{\rm SSF}(M_\psi^2/M_\phi^2) \bigr]$ & $\delta a_{\mu} < 0$ \\
\hline
$\phi_{T} + \psi_{A}$ & Fig.~\ref{fig:twop}~(a,b,c,d) & 
{$\frac{m_\mu^2Y^2}{16\pi^2M_\phi^2}3 F_{\rm SSF}(M_\psi^2/M_\phi^2)$} & $\delta a_{\mu} < 0$ \\
\hline
$V^0 + \psi^\pm$ & Fig.~\ref{fig:twop}~(g) &
$\frac{m_\mu^2g^2}{16\pi^2M_V^2}F_{\rm FFV}(M_\psi^2/M_V^2)$ & $\delta a_{\mu} < 0$ \\
\hline
$V^\pm + \psi^0$ & Fig.~\ref{fig:twop}~(h) &
$\frac{m_\mu^2g^2}{16\pi^2M_V^2}F_{\rm VVF}(M_\psi^2/M_V^2)$ & $\delta a_{\mu} > 0$ \\
\hline
$V_A + \psi_D$ & Fig.~\ref{fig:twop}~(g,h) &
$\frac{m_\mu^2g^2}{64\pi^2M_V^2}\bigl[F_{\rm FFV}(M_\psi^2/M_V^2)
  + 2F_{\rm VVF}(M_\psi^2/M_V^2) \bigr]$ & $\delta a_{\mu} > 0$ \\
\hline
\end{tabular}
\caption{Correction $\delta a_\mu$ from two new fields with different spin. The functions $F_{\rm XYZ}$ and $G_{\rm XYZ}$ are defined in (\ref{eq:loopfunctions}). The notation follows the one introduced in Section~\ref{sec:twofields}.}
\label{tab:twop}
\end{table}

Contributions to $a_{\mu}$ of two mixing vector fermions and SM bosons in the loop can be expressed as
\begin{align}
a_{\mu}^Z(F) &= \frac{G_F}{2\sqrt{2}\pi^2}\biggl[m_{\mu}^2\bigl((g_L^{ZF})^2+(g_R^{ZF})^2\bigr)F_{\rm FFV}\left(\frac{M_F^2}{M_Z^2}\right)
+ m_{\mu}M_F\,g_L^{ZF}g_R^{ZF}\,G_{\rm FFV}\left(\frac{M_F^2}{M_Z^2}\right)\biggr]\,,\nonumber\\[.5ex]
a_{\mu}^W(N) &= \frac{G_F}{4\sqrt{2}\pi^2}\biggl[m_{\mu}^2\big((g_L^{WN})^2+(g_R^{WN})^2\big)F_{\rm VVF}\left(\frac{M_N^2}{M_W^2}\right)
+ m_{\mu}M_N\,g_L^{WN}g_R^{WN}\,G_{\rm VVF}\left(\frac{M_N^2}{M_W^2}\right)\biggr]\,,\nonumber\\[.5ex]
a_{\mu}^H(F) &= \frac{G_F}{16\sqrt{2}\pi^2}\biggl[m_{\mu}^2\big((g_L^{HF})^2+(g_R^{HF})^2\big)F_{\rm FFS}\left(\frac{M_F^2}{M_H^2}\right)
+ m_{\mu}M_F\,g_L^{HF}g_R^{HF}\,G_{\rm FFS}\left(\frac{M_F^2}{M_H^2}\right)\biggr]\,,
\label{eq:amuWZH}
\end{align}
where $F=\mu^-,\psi^-,\psi_D^-,\psi_A^-,\psi_T^-$ and $N=\psi^0,\psi_D^0,\psi_A^0,\psi_T^0$. The contributions of doubly-charged fermions are given by
\begin{align}\label{eq:amuWCC}
a_{\mu}^W(C) &= \frac{G_F}{4\sqrt{2}\pi^2}\big[m_{\mu}^2\big((g_L^{WC})^2+(g_R^{WC})^2\big)\bigl\{2F_{\rm FFV}\left(\frac{M_C^2}{M_W^2}\right) - F_{\rm VVF}\left(\frac{M_C^2}{M_W^2}\right) \bigr\}\\\nonumber
 & \qquad\qquad\qquad\qquad\qquad + m_{\mu}M_C\,g_L^{WC}g_R^{WC}\bigl\{2G_{\rm FFV}\left(\frac{M_C^2}{M_W^2}\right) - G_{\rm VVF}\left(\frac{M_C^2}{M_W^2}\right) \bigr\}\big]\,,
\end{align}
with $C=\psi_T^{--}$. The couplings $g_{L,R}^{BF}$ of new vector fermions to muons and SM bosons (as induced by electroweak symmetry breaking) are defined as
\begin{align}
\mathcal{L} &\supset \frac{g}{\sqrt{2}}\,g_{L,R}^{WN}\,W^+_{\mu}\overline{N}\gamma^{\mu}\mu^-_{L,R} + \frac{g}{\sqrt{2}}\,g_{L,R}^{WC}\,W^-_{\mu}\overline{C}\gamma^{\mu}\mu^-_{L,R}\\\nonumber
            &\qquad\qquad\qquad\qquad + \frac{g}{c_W}\,g_{L,R}^{ZF}\,Z_{\mu}\overline{F}\gamma^{\mu}\mu^-_{L,R} - \frac{1}{\sqrt{2}}\,g_{L,R}^{HF}\,\overline{F}\mu^-_{L,R} + \rm{h.c.}\,.
\end{align}
For the different scenarios considered in this work, they are listed in Tables~\ref{tab:mixedDS} and \ref{tab:mixedDA}. We have expanded these couplings in terms of the mixing parameters $\epsilon_i = Y_iv/M_i$ and \mbox{$\omega_{ij} = Y_{ij}v/(M_i-M_j)$}, with $i=S,N,D,A,T$ and $ij=SD,DS$ etc. The respective Yukawa couplings are defined in (\ref{fn})--(\ref{ft}) and (\ref{eq:DSmixing})--(\ref{eq:DTmixing}). Our results agree with \cite{Kannike:2011ng} for the case $\psi_D+\psi^\pm$. However, we find a different sign in front of the contribution with one doubly-charged fermion and two $W$ bosons in the loop with respect to the one in (3.20) and (3.21) in \cite{Kannike:2011ng}.\\
\vspace*{0.7cm}

\begin{table}[!ht]
\centering
\renewcommand{\arraystretch}{1.5}
\begin{tabular}{|l|c|c|}
\hline
$\psi_D+\psi^{\pm}$ & $g_L^{BF}$ & $g_R^{BF}$\\
\hline
$Z\overline{\mu^-}\mu^-$ & $-\frac{1}{2}+s_W^2+\frac{\epsilon_S^2}{4}$ & $s_W^2 -\frac{\epsilon_D^2}{4}$\\
$Z\overline{\psi_D^-}\mu^-$ & $\frac{M_D\omega_{SD} - M_S\omega_{DS}}{4(M_S+M_D)}\epsilon_S$ & $\frac{\epsilon_D}{2\sqrt{2}}$ \\
$Z\overline{\psi^-}\mu^-$ & $-\frac{\epsilon_S}{2\sqrt{2}}$ & $\frac{M_S\omega_{SD} - M_D\omega_{DS}}{4(M_S+M_D)}\epsilon_D$ \\
\hline
$W^+\overline{\nu}\mu^-$ & $1 -\frac{\epsilon_S^2}{4}$ & $0$ \\
$W^+\overline{\psi_D^0}\mu^-$ & $ -\frac{m_{\mu}}{M_D}\frac{\epsilon_D}{\sqrt{2}} + \frac{(M_S-M_D)\omega_{SD}}{2M_D}\epsilon_S$  & $-\frac{\epsilon_D}{\sqrt{2}}$ \\
\hline
$H\overline{\mu^-}\mu^-$ & $\sqrt{2}\frac{m_{\mu}}{M_H}$ & $\sqrt{2}\frac{m_{\mu}}{M_H}$ \\
$H\overline{\psi_D^-}\mu^-$ & $\frac{m_{\mu}}{M_H}\epsilon_D + \frac{(M_D^2-2M_S^2)\omega_{SD} + M_SM_D\omega_{DS}}{\sqrt{2}M_H(M_S+M_D)}\epsilon_S$  & $\frac{M_D}{M_H}\epsilon_D$ \\
$H\overline{\psi^-}\mu^-$ & $\frac{M_S}{M_H}\epsilon_S$ & $\frac{m_{\mu}}{M_H}\epsilon_S + \frac{(2M_D^2-M_S^2)\omega_{SD} - M_SM_D\omega_{DS}}{\sqrt{2}M_H(M_S+M_D)}\epsilon_D$ \\
\hline
\hline
$\psi_D+\psi^0$ & $g_L^{BF}$ & $g_R^{BF}$ \\
\hline
$Z\overline{\mu^-}\mu^-$ & $-\frac{1}{2}+s_W^2$ & $s_W^2 -\frac{\epsilon_D^2}{4}$\\
$Z\overline{\psi_D^-}\mu^-$ & $0$ & $\frac{\epsilon_D}{2\sqrt{2}}$ \\
\hline
$W^+\overline{\nu}\mu^-$ & $1 -\frac{\epsilon_{N}^2}{4}$ & $0$ \\
$W^+\overline{\psi_D^0}\mu^-$ & $ -\frac{m_{\mu}}{M_D}\frac{\epsilon_D}{\sqrt{2}} + \frac{M_N(M_D\omega_{DN} - M_S\omega_{ND})}{2M_D(M_D+M_N)}\epsilon_{N}$ & $-\frac{\epsilon_D}{\sqrt{2}}$\\
$W^+\overline{\psi^0}\mu^-$ & $\frac{\epsilon_N}{\sqrt{2}}$ & $\frac{M_D\omega_{DN} - M_S\omega_{ND}}{2(M_D+M_N)}\epsilon_D$ \\
\hline
$H\overline{\mu^-}\mu^-$ & $\sqrt{2}\frac{m_{\mu}}{M_H}$ & $\sqrt{2}\frac{m_{\mu}}{M_H}$ \\
$H\overline{\psi_D^-}\mu^-$ & $\frac{m_{\mu}}{M_H}\epsilon_D$ & $\frac{M_D}{M_H}\epsilon_D$ \\
\hline
\end{tabular}
\caption{Couplings $g_{L,R}^{BF}$ of a new fermion $F$ and a SM boson $B$ to a left- or right-handed muon in scenarios with mixing vector fermion doublet and singlet in the vertex loop.}
\label{tab:mixedDS}
\end{table}
\begin{table}[!p]
\centering
\renewcommand{\arraystretch}{1.5}
\begin{tabular}{|l|c|c|}
\hline
$\psi_D+\psi_A$ & $g_L^{BF}$ & $g_R^{BF}$ \\
\hline
$Z\overline{\mu^-}\mu^-$  & $-\frac{1}{2} + s_W^2 -\frac{\epsilon_A^2}{4}$ & $s_W^2 -\frac{\epsilon_D^2}{4}$ \\
$Z\overline{\psi_D^-}\mu^-$ & $\frac{M_A\omega_{DA} - M_D\omega_{AD}}{4(M_D+M_A)}\epsilon_A$ & $\frac{\epsilon_D}{2\sqrt{2}}$ \\
$Z\overline{\psi_A^-}\mu^-$ & $\frac{\epsilon_A}{2\sqrt{2}}$ & $\frac{m_{\mu}}{M_A}\frac{\epsilon_A}{\sqrt{2}} + \frac{(2M_D^2-M_A^2)\omega_{AD} - M_AM_D\omega_{DA}}{4M_A(M_D+M_A)}\epsilon_D$ \\
\hline
$W^+\overline{\nu}\mu^-$ & $1 + \frac{\epsilon_A^2}{8}$ & $0$ \\
$W^+\overline{\psi_D^0}\mu^-$ & $-\frac{m_{\mu}}{M_D}\frac{\epsilon_D}{\sqrt{2}} + \frac{M_A(M_A\omega_{AD} - M_D\omega_{DA})}{4M_D(M_D+M_A)}\epsilon_A$ & $-\frac{\epsilon_D}{\sqrt{2}}$ \\
$W^+\overline{\psi_A^0}\mu^-$ & $-\frac{\epsilon_A}{2}$ & $ -\frac{m_{\mu}}{M_A}\epsilon_A - \frac{(2M_D^2-M_A^2)\omega_{AD} - M_AM_D\omega_{DA}}{2\sqrt{2}M_A(M_D+M_A)} \epsilon_D $ \\
\hline
$H\overline{\mu^-}\mu^-$ & $\sqrt{2}\frac{m_{\mu}}{M_H}$ & $\sqrt{2}\frac{m_{\mu}}{M_H}$ \\
$H\overline{\psi_D^-}\mu^-$ & $\frac{m_{\mu}}{M_H}\epsilon_D + \frac{(M_D^2-2M_A^2)\omega_{AD} + M_AM_D\omega_{DA}}{\sqrt{2}M_H(M_D+M_A)} \epsilon_A$ & $\frac{M_D}{M_H}\epsilon_D$ \\
$H\overline{\psi_A^-}\mu^-$ & $\frac{M_A}{M_H}\epsilon_A$ & $\frac{m_{\mu}}{M_H}\epsilon_A + \frac{(2M_D^2-M_A^2)\omega_{AD} - M_AM_D\omega_{DA}}{\sqrt{2}M_H(M_D+M_A)}\epsilon_D$ \\
\hline
\hline
$\psi_D+\psi_T$ & $g_L^{BF}$ & $g_R^{BF}$ \\
\hline
$Z\overline{\mu^-}\mu^-$  & $-\frac{1}{2} + s_W^2 +\frac{\epsilon_T^2}{8}$ & $s_W^2 -\frac{\epsilon_D^2}{4}$ \\
$Z\overline{\psi_D^-}\mu^-$ & $\frac{M_D\omega_{TD} - M_T\omega_{DT}}{8(M_D+M_T)}\epsilon_T$ & $\frac{\epsilon_D}{2\sqrt{2}}$ \\
$Z\overline{\psi_T^-}\mu^-$ & $\frac{\epsilon_T}{4}$ & $\frac{M_D\omega_{DT} - M_T\omega_{TD}}{4\sqrt{2}(M_D+M_T)}\epsilon_D$ \\
\hline
$W^+\overline{\nu}\mu^-$ & $1 - \frac{7\epsilon_T^2}{8}$ & $0$ \\
$W^+\overline{\psi_D^0}\mu^-$ & $-\frac{m_{\mu}}{M_D}\frac{\epsilon_D}{\sqrt{2}} - \frac{(3M_D^2+M_T^2)\omega_{TD} - 4M_TM_D\omega_{DT}}{4M_D(M_D+M_T)}\epsilon_T$ & $-\frac{\epsilon_D}{\sqrt{2}}$ \\
$W^+\overline{\psi_T^0}\mu^-$ & $\sqrt{2}\epsilon_T$ & $\frac{m_{\mu}}{M_T}\frac{\epsilon_T}{\sqrt{2}} - \frac{(2M_T^2-M_D^2)\omega_{TD} - M_TM_D\omega_{DT}}{2M_T(M_D+M_T)}\epsilon_D$ \\
$W^+\overline{\psi_T^{--}}\mu^-$ & $-\frac{\epsilon_T}{\sqrt{2}}$ & $-\frac{m_{\mu}}{M_T}\frac{\epsilon_T}{\sqrt{2}} -\frac{(M_D-M_T)\omega_{TD}}{2M_T}\epsilon_D$ \\
\hline
$H\overline{\mu^-}\mu^-$ & $\sqrt{2}\frac{m_{\mu}}{M_H}$ & $\sqrt{2}\frac{m_{\mu}}{M_H}$ \\
$H\overline{\psi_D^-}\mu^-$ & $\frac{m_{\mu}}{M_H}\epsilon_D + \frac{(M_D^2-2M_T^2)\omega_{TD} + M_TM_D\omega_{DT}}{2\sqrt{2}M_H(M_D+M_T)} \epsilon_T$ & $\frac{M_D}{M_H}\epsilon_D$ \\
$H\overline{\psi_T^-}\mu^-$ & $-\frac{M_T}{M_H}\frac{\epsilon_T}{\sqrt{2}}$ & $ -\frac{m_{\mu}}{M_H}\frac{\epsilon_T}{\sqrt{2}} - \frac{(2M_D^2-M_T^2)\omega_{TD} - M_TM_D\omega_{DT}}{2M_H(M_D+M_T)}\epsilon_D$ \\
\hline
\end{tabular}
\caption{Couplings $g_{L,R}^{BF}$ of a new fermion $F$ and a SM boson $B$ to a left- or right-handed muon in scenarios with mixing vector fermion doublet and triplet in the vertex loop.}
\label{tab:mixedDA}
\end{table}


\section{Four-lepton contact interactions}\label{sec:4lci}
Four-lepton interactions are generated at the one-loop level by two new fields with different spin. The results for all combinations of fields defined in Table~\ref{tab:notation} that yield a positive contribution $\delta a_{\mu}$ are listed in Table~\ref{tab:4lci}. The corresponding loop functions read
\begin{align}\label{eq:ciloop}
F_{\rm FS}(x) &= \tfrac{1}{(x-1)^3} \bigl [x^2-1-2x\ln x \bigr],\\\nonumber
F_{\rm FV}(x) &= \tfrac{1}{(x-1)^3} \bigl [x^4-16x^3+19x^2+2(3x^2+4x-4)x\ln x-4\bigr].
\end{align}
Notice that these results are model-independent and applicable to any scenario with couplings of two new fields to leptons.

\begin{table}[!h]
\centering
\renewcommand{\arraystretch}{1.5}
\begin{tabular}{|l|c|c|}
\hline
$\phi^0 + \psi^\pm$ & -- & $0$ \\
\hline
$\phi_{D} + \psi_{D}$ & Fig.~\ref{fig:box} (a,b) & $\frac{Y^4}{32\pi^2M_\phi^2} F_{\rm FS}(M_\psi^2/M_\phi^2)\,\mathcal{O}_{\rm RR}$ \\
\hline
$\phi_{D} + \psi_{A}$ & Fig.~\ref{fig:box} (a,b) & $\frac{5Y^4}{256\pi^2M_\phi^2} F_{\rm FS}(M_\psi^2/M_\phi^2)\,\mathcal{O}_{\rm LL}$ \\ 
\hline
$\phi_{D} + \psi_{T}$ & Fig.~\ref{fig:box} (a,c) & $\frac{5Y^4}{256\pi^2M_\phi^2} F_{\rm FS}(M_\psi^2/M_\phi^2)\,\mathcal{O}_{\rm LL}$ \\
\hline
$\phi_{A} + \psi_{T}$ & -- & $0$ \\
\hline
$V^\pm + \psi^0$ & Fig.~\ref{fig:box} (d) & $\frac{g^4}{64\pi^2M_V^2} F_{\rm FV}(M_\psi^2/M_V^2)\,\mathcal{O}_{\rm RR}$ \\
\hline
$V_A + \psi_D$ & Fig.~\ref{fig:box} (d,e) & $\frac{g^4}{256\pi^2M_V^2}\bigl[F_{\rm FV}(M_\psi^2/M_V^2) - 3F_{\rm FS}(M_\psi^2/M_V^2)\bigr]\,\mathcal{O}_{\rm LL}$ \\
\hline
\end{tabular}
\caption{Effective four-lepton interactions $\mathcal{C}_{AA}\mathcal{O}_{AA}$ for pairs of new fields leading to $\delta a_{\mu} > 0$. The loop functions $F_{\rm FS}$ and $F_{\rm FV}$ are defined in (\ref{eq:ciloop}). The notation has been introduced in Section~\ref{sec:twofields}.}
\label{tab:4lci}
\end{table}



\end{document}